\documentclass[fleqn,usenatbib]{mnras}

\usepackage{newtxtext,newtxmath}

\usepackage[T1]{fontenc}

\DeclareRobustCommand{\VAN}[3]{#2}
\let\VANthebibliography\thebibliography
\def\thebibliography{\DeclareRobustCommand{\VAN}[3]{##3}\VANthebibliography}

\usepackage{subfig}

\usepackage{graphicx}	
\usepackage{amsmath}	
\usepackage{xcolor}
\usepackage{units}      
\definecolor{lime}{HTML}{A6CE39}
\usepackage{multirow}

\newcommand{\rev}[1]{{{#1}}}

\title[Impact behaviours of three planet systems]{Orbital stability of compact three-planet systems, II: Post-instability impact behaviour}

\author[P. Bartram et al.]{
Peter Bartram,$^{ 1}$\thanks{E-mail: p.bartram@soton.ac.uk} 
Alexander Wittig,$^{1}$ 
Jack J. Lissauer,$^{2}$ 
Sacha Gavino$^{3}$
and Hodei Urrutxua$^{4}$
\\
$^{1}$University of Southampton, Southampton, UK\\    
$^{2}$Space Science \& Astrobiology Division, MS 245-3, NASA Ames Research Center, Moffett Field, CA 94035, USA \\
$^{3}$Laboratoire d’Astrophysique de Bordeaux, Universit\'e de Bordeaux, CNRS, F-33615, Pessac, France \\
$^{4}$European Institute for Aviation Training and Accreditation, Universidad
Rey Juan Carlos, Madrid, Spain
}

\date{Accepted XXX. Received YYY; in original form ZZZ}

\pubyear{2015}

\begin{document}
\label{firstpage}
\pagerange{\pageref{firstpage}--\pageref{lastpage}}
\maketitle

\begin{abstract}
Recent observational missions have uncovered a significant number of compact multi-exoplanet systems. The tight orbital spacing of these systems has led to much effort being applied to the
understanding of their stability; however, a key limitation of the majority of these studies is the termination of simulations as soon as the orbits of two planets cross. In this work we explore the stability of compact, three-planet systems and continue our simulations all the way to the first collision of planets
to yield a better understanding of the lifetime of these systems.
We perform over $25,000$ integrations of a Sun-like star orbited by three Earth-like secondaries for up to a billion orbits to explore a wide parameter space of initial conditions in both the co-planar and inclined cases, with a focus on the initial orbital spacing. We calculate the probability of collision over time and determine the probability of collision between specific pairs of planets. We find systems that persist for over $10^8$ orbits after an orbital crossing and show how the post-instability survival time of systems depends upon the initial orbital separation, mutual inclination, planetary radius, and the closest encounter experienced. Additionally, we examine the effects of very small changes in the initial positions of the planets upon the time to collision and show the effect that the choice of integrator can have upon simulation results. We generalise our results throughout to show both the behaviour of systems with an inner planet initially located at $1$ AU and $0.25$ AU.

\end{abstract}




\section{Introduction}

The now retired NASA Kepler Space Telescope is responsible for observations leading to the confirmation of hundreds of multi-planet systems \citep{Lissauer2014, Rowe2014}. Of these systems, as many as six percent are thought to be compact \citep{Wu2019}, containing planets that are much more closely spaced than the inner planets of our own Solar System. These discoveries have naturally led to many questions being asked about the long-term stability of compact exoplanet systems. Indeed, it is even possible that compact planetary embryos existed interior to Venus's current orbit that have subsequently been \rev{expelled} from this region due to orbital instabilities \citep{Volk2015}. Within the class of observed compact systems, a large population of planets have been observed with a mass \citep{Mayor2011} and radius \citep{Petigura2013} between that of Earth and Neptune. Moreover, the observed orbital architecture is such that mutual inclinations are small, typically in the region of $1^{\circ}$ to $2^{\circ}$ \citep{Fabrycky2014}, while eccentricities are also found to be small\rev{, on average $\Bar{e} \approx 0.04$} \citep{Xie2016}. An archetypal example of these systems, albeit containing six planets, is Kepler-11 \citep{Lissauer2011}.
Exoplanet systems with orbital spacings much greater than that required for stability are also present in the Kepler data set. It is \rev{a favoured hypothesis} that this orbital architecture is a result of dynamical instabilities in much more compact systems leading to close encounters and orbital reconfiguration \citep{Pu2015}. Understanding of the stability and evolution of compact exoplanet systems is therefore not only important for making sense of observations but also for understanding the planetary formation process as a whole.

Characterisation of the stability of three or more planet systems can be approached in several ways. Analytical models have been built that can predict the lifetime of three planet systems based upon resonance overlap \citep{Wisdom1980, Quillen2011, Petit2020}. Recently, machine learning approaches have also been developed that, after being guided by a training set of $10^9$ year integrations, can use far shorter integrations to predict with surprisingly high accuracy which given exoplanet systems will remain stable for a billion orbital period\rev{s} \citep{Tamayo2016, Tamayo2020}.
However, the most common approach to the problem, and the one employed in this paper, is the use of n-body simulation \citep{Chambers1999, Smith2009, Obertas2017, Hussain2020, Lissauer2021}.

The majority of studies performed take a subset of the possible input parameter space for a compact, near-circular, near co-planar system of a given number of planets and then evolve this system forward in time checking for either the first close approach, typically specified as one Hill radius, $r_H$, or waiting for an orbital crossing to occur: this is then termed the instability event.
Throughout this work we will use orbital crossing as our definition of an instability event and refer to the time at this point as the crossing time. \citet{Rice2018} found that systems containing four Neptune-size \rev{and Neptune-mass} planets \rev{initially located at $1$~AU} can continue to evolve after an instability event for over ten million dynamical periods before a collision of planets, meaning that the commonly used instability metric may not capture the entire evolution of the system. Given that the manner in which these planets collide determines the final orbital architecture it is important properly to understand this phase of the exoplanet system life cycle.

Our study builds upon the work done by \citet{Rice2018} and \citet{Lissauer2021} by considering the post-instability evolution of compact, Earth-analogue, three-planet systems across a large range of initial orbital separations equally spaced in units of mutual Hill radii. We create three integration suites called the standard suite, perturbed suite and inclined suite, and perform $4,835$ integrations each in the first two and a further $11,200$ in the final one. We continue integrations up until the time of first collision between planets or for $10^8$ or $10^9$ orbits depending on the experiment.

In section \ref{sec: simulations} of this paper we describe the methodology used for our integrations including the initial conditions for each integration suite, the integration packages used and the termination criteria. Section \ref{sec: standard suite results} contains the results of all standard suite integrations: section \ref{sec: time scale to planet planet collision} details the time scales for orbital crossing and collision between pairs of planets, and details collision probabilities over time for various initial configurations of systems; the effects of small changes in initial orbital longitudes upon these results are then examined in section \ref{sec: sensitivity to initial conditions}; and, finally, section \ref{sec: which planets collide} examines the probabilities of particular pairs of planets colliding.
Section \ref{sec: inclined integration suite} introduces the results of inclined suite integrations: in section \ref{sec: dynamic heating} we explore the heating of what are initially dynamically cold systems that eventually enables orbital crossing and collision; here, we find that the three-planet Earth-mass systems behave in a similar manner to the four-planet Neptune-mass case but follow a different power law. Section \ref{sec: inclined time scale to planet planet collision} examines the time scales leading to collision in the inclined case and shows that the survival time after crossing can be a non-trivial fraction of the main-sequence lifetime of stars. In addition, this section also looks at the effects on lifetime of systems dependent on the distance from the innermost planet to the star and the initial inclination. We  summarise our findings in section \ref{sec: conclusions}.

\section{Methods}
\label{sec: simulations}
We have chosen to simulate three-planet systems comprising of analogues from our own Solar System. The central body in each of our systems is a one solar mass star, $m_0$  = 1 $\textrm{M}_\odot$. Each of the planets within the systems are Earth mass, $m_j$ = 1 $\textrm{M}_\oplus$ where $j={1,2,3}$ with a planetary radius also equal to that of Earth, $\textrm{R}_p = \textrm{R}_\oplus$. Planets are placed on initially circular orbits orbiting the star in a common direction with the innermost planet located at $1~\textrm{AU}$. Time throughout this work is provided in units of initial orbital period of the innermost planet, this means that the crossing time is invariant to rescaling of the system so long as initial orbital period ratios between bodies are maintained along with mass-ratios of planets and star.

\subsection{Initial semi-major axes}
Initial semi-major axes $a_j$ of systems are evenly spaced in terms of mutual Hill radii. The mutual Hill radii \rev{are defined as}

\begin{equation}
r_{H_{j, j+1}} = \left( \dfrac{m_j + m_{j+1}}{m_0 + \sum^{j-1}_{k=1} m_k} \right)^{\frac{1}{3}} \left( \dfrac{a_{j} + a_{j+1}}{2}\right).
\end{equation}

This allows for a dimensionless value $\beta$ to be defined to specify the even spacing of adjacent planetary orbits in units of their mutual Hill radii as

\begin{equation}
    \beta \equiv \dfrac{a_{j+1} - a_j}{r_{H_{j, j+1}}}.
\end{equation}

Therefore, the initial semi-major axes of adjacent planets are chosen to be such that 
\begin{equation}
    \begin{split}
    &a_{j+1} = a_j + \beta r_{H_{j, j+1}} \\
    &= a_j\left[1 + \dfrac{\beta}{2} \left( \dfrac{m_j + m_{j+1}}{m_0 + \sum^{j-1}_{k=1} m_k} \right)^{\frac{1}{3}} \right]\left[1 - \dfrac{\beta}{2} \left( \dfrac{m_j + m_{j+1}}{m_0 + \sum^{j-1}_{k=1} m_k} \right)^{\frac{1}{3}} \right]^{-1}.    
    \end{split}
    \label{eq: semimajor axis spacing}
\end{equation}
    
 The innermost planet is placed such that it has a semi-major axis of $1\, \textrm{AU}$, and all other semi-major axes are chosen through Eq.~\eqref{eq: semimajor axis spacing}. We refer to this configuration as a \emph{system at $1$ AU}. Likewise, later on, when results are generalised to include systems with an innermost planet located at $0.25$ AU with other planets spaced as per Eq.~\eqref{eq: semimajor axis spacing} we refer to it as a \emph{system at $0.25$ AU}.

\begin{figure*}
    \centering
    \includegraphics[width=0.975\textwidth]{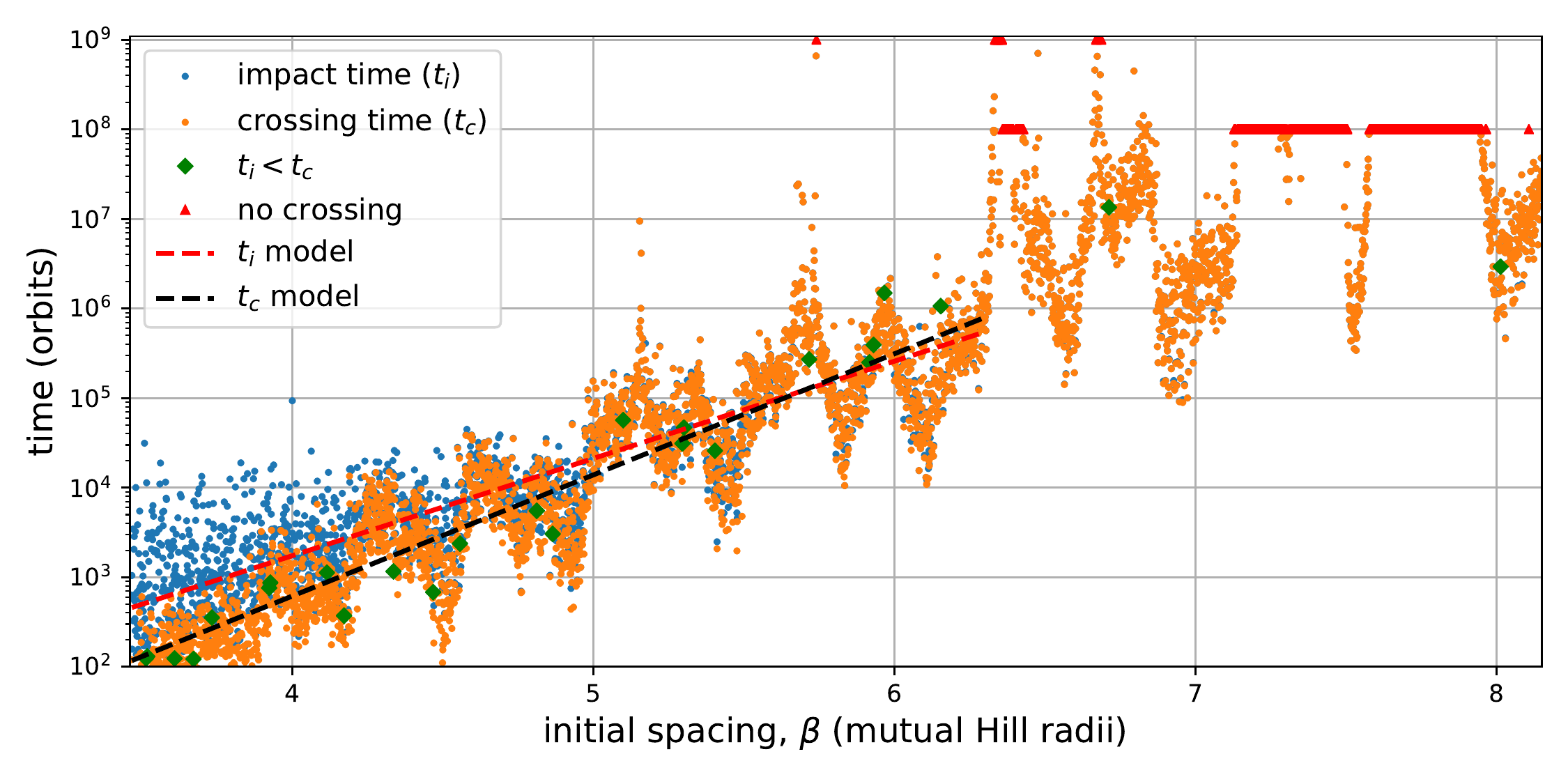}
    \caption{Plot showing the crossing time, $t_c,$ and impact time, $t_i$, for all integrations in the standard suite \rev{for systems at $1$~AU}. Simulations are run for up to $10^9$ orbits in general but some are terminate at $10^8$ orbits to save on computation. Orbits are specified by the initial period of the innermost planet. Impacts that take place before a crossing are highlighted \rev{by a} green \rev{diamond} whereas systems that did not cross within the maximum simulation time are marked with a red triangle. \rev{Models fitted to the crossing and impact times according to Eq.~\ref{eq: model} are shown as a dashed black and a dashed red line, respectively.}}   
    \label{fig: orbital crossing at 1AU}
\end{figure*}

 \subsection{Stopping criteria and integration packages}
We have opted to use the Terrestrial Exoplanet Simulator (TES) \footnote{Code available at \url{https://github.com/PeterBartram/TES}}  package  to perform our integrations \citep{Bartram2021a}. TES is a new numerical integration package written in C++ for propagating exoplanet systems. This package combines an integrator that follows Brouwer's law \citep{Brouwer1937} with a new special perturbation method to allow for reduced run-times and decreased numerical error resulting in, e.g., improved energy conservation. Additionally, this tool has been designed to allow for integration all the way to collision of terrestrial mass planets to machine precision. TES can be run using C++ directly, or through a python interface allowing for ease of use and for multiple integrations to be performed in parallel. Throughout our simulations we have opted to use TES with a non-dimensional tolerance of $1 \times 10^{-8}$ which has ensured that the relative energy error in all simulations, even after collision, and for the longest lived systems, is maintained below $1 \times 10^{-13}$.
To validate our own results, we also repeated all of our standard suite integrations making use of \rev{IAS15} \citep{Rein2014} within the REBOUND package \citep{Rein2012}. The results from this comparison can be found in Appendix \ref{appendix: integrator comparison}.

As mentioned before, time is measured  by periods of the innermost planet in the system throughout this work, meaning that all times are specified in units of orbits or dynamical periods.
Integrations run until either a collision is detected or the simulation reaches a maximum time of $10^8$ or $10^9$ dynamical periods, depending on the experiment.

In order to detect an orbital crossing, the orbital elements of each planet are calculated at every step within each integration. These are then compared to determine the time at which the apoapsis of a planet crosses the periapsis of the exterior adjacent planet. We define this as the \emph{crossing time} and denote it $t_c$. 
Moreover, also at each step, the mutual separations of each of the planets are calculated so that collisions can be detected. The metric of two planets coming within $2 \textrm{R}_\oplus$ of one another is used for collision detection. We define the time at which this occurs as the \emph{impact time} and denote it $t_i$. We also define the \emph{post-crossing survival time}, $t_s$, of a system to be the time that the system persists without a collision after the point of orbital crossing: $$t_s \equiv t_i - t_c.$$
All encounters closer than any experienced previously are recorded such that it is possible to generalise the collision results to systems with planetary radii greater than that of the Earth or, equivalently, initial orbital radii closer than $1~\textrm{AU}$. \rev{We use this generalisation to consider systems at $0.25$~AU and $1$~AU for all integration suites.} We also define the time of closest encounter prior to collision as the \emph{closest encounter time}, $t_e$.
To ensure bit-wise identical initial conditions as in \citet{Lissauer2021}, initial conditions are specified as orbital elements which are then entered in to the MERCURY \citep{Chambers1999} integration package in order to generate an initial state vector which is then provided to either TES or REBOUND. Table~\ref{tab: time symbols} contains a summary of all symbols related to simulations event times.

\begin{table}
	\centering
	\caption{Summary of all simulation event time symbols used.}
	\label{tab: time symbols}
	\begin{tabular}{lcc} 
		\hline
		Symbol & Definition \\
         \hline
        $t_c$ & crossing time    \\
        $t_i$ & impact time    \\
        $t_s$ & post-crossing survival time    \\
        $t_e$ & closest encounter time    \\
        \hline
	\end{tabular}
\end{table}

 \subsection{Standard integration suite}
 \label{sec: standard suite}
 The first suite of integrations is composed of $4,835$ orbital configurations and is termed our standard suite. In this suite systems are on initially circular, co-planar orbits with an initial mean anomaly for the $j_{\textrm{th}}$ planet $M_j = 2 \pi j \lambda$ radians where $\lambda \equiv \dfrac{1}{2}\left( 1+ \sqrt{5}\right)$, i.e., the golden ratio, and are merely chosen to avoid special orientations.
 As we wish to study the effects of the initial spacing of planets upon impact timescales we choose a high resolution in $\beta$ such that there are $1 \times 10^{3}$ integrations per unit $\beta$ over the range $\beta = [3.\rev{46}5, 8.3]$. Generally, integrations are terminated after $10^\rev{9}$ orbits if a collision is not encountered. However, in certain areas we have chosen to \rev{limit} integrations to $10^\rev{8}$ orbits \rev{to save on computation}; these regions are clearly marked on any plots.

\subsection{Perturbed integration suite} 
\label{sec: perturbed suite}
The second integration suite is termed our perturbed suite and is also composed of $4,835$ integrations. The only difference between the initial conditions of the standard suite and the perturbed suite is that in the latter case the innermost planet is perturbed by $100~\textrm{m}$ along its orbital arc. We strictly terminate integration at $1 \times 10^{8}$ orbital period\rev{s} of the innermost planet in this suite. This suite is used to examine the effects of very small changes in initial conditions upon crossing and impact time.

\subsection{Inclined integration suite}
\label{sec: inclined suite}
The final integration suite is the inclined suite and is composed of \rev{$16,800$} integrations. \rev{Of these, $15$ did not complete in the available CPU time and have been excluded from the dataset. This is equivalent to $0.09\%$ of inclined integrations, and we therefore do not believe this will have biased the dataset in any meaningful way.}
We choose initial conditions across a subset of the available parameter space manually rather than randomly and perform integrations for a maximum simulation time of $1 \times 10^{8}$ orbital periods of the innermost planet. To make best use of computational resources we limit this study to the range $\beta = [3.5, 6.3]$ and perform experiments uniformly spaced in $\beta$ with fifty values per unit $\beta$. At each value of $\beta$ we perform one hundred and twenty experiments where the initial values of semi-major axis, eccentricity and mean longitude are the same as in the standard suite. 

Planets are, however, inclined relative to each other in one of four ways: one of inner, middle or outer planet inclined above the orbital plane of the system, and also with the middle planet above and the outer planet below. For each such configuration of relative inclination fifteen initial values of inclination are logarithmically spaced between $i_0 = 0.06^\circ$  and $i_0 = 0.58^\circ$, yielding an initial orbital height ranging from $0.10 ~ r_H$ to $r_H$. The distribution of initial inclinations within this range is such that ten values are used between $i_0 = 0.24^\circ$ and $i_0 = 0.58^\circ$ and five values are used over the region $i_0 = 0.06^\circ$ and $i_0 = 0.24^\circ$.
Finally, two values are chosen for the ascending nodes $\Omega$: either according to the golden ratio in Section \ref{sec: simulations} such that $M_j = \Omega_j$ or equally spaced such that $\Omega_j = [0^\circ, 120^\circ, 240^\circ]$. 

The full state vector of each simulation is output to file once every ten thousand orbital periods; additionally, each planetary flyby closer than any other previously observed is also recorded. 

\begin{figure}
    \centering
    \includegraphics[width=0.475\textwidth]{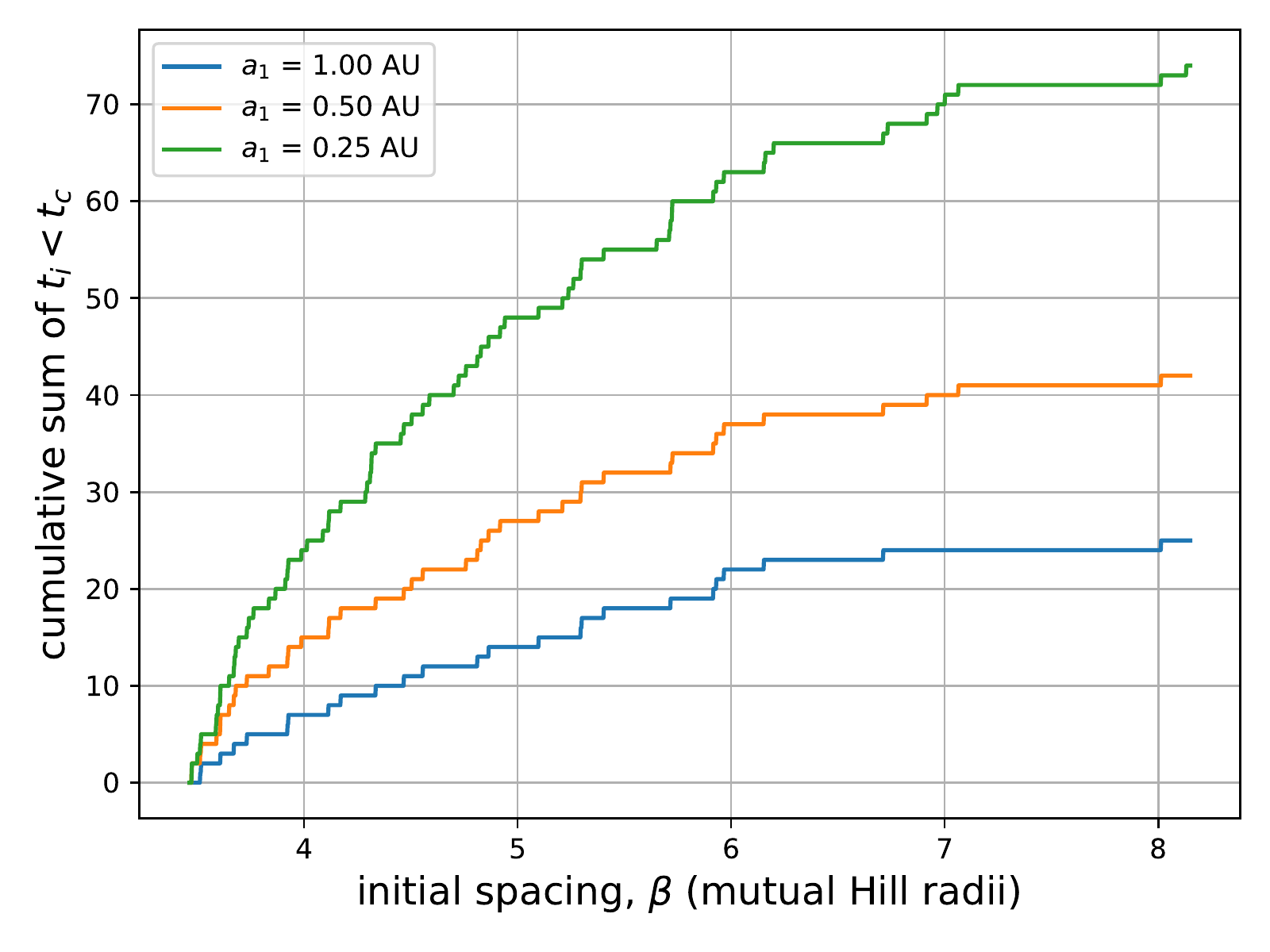}
    \caption{C\rev{umulative sum of integrations with a collision before orbital crossing} for various initial values for semi-major axis of the innermost planet. The flat region between beta $\beta = 7$ and $\beta = 8$ is due to systems not experiencing an orbital crossing within the maximum simulation time \rev{ in that region and integrations being terminated early (see the red triangles in Figure~\ref{fig: orbital crossing at 1AU}). A large fraction of integrations with an initial spacing $\beta > 6.3$ were stopped at $10^8$ orbits so results beyond this value cannot be considered to have been drawn from a uniform sample.}}
    \label{fig: cumsum ti before tc}
\end{figure}

\begin{figure*}
    \centering
    \includegraphics[width=0.975\textwidth]{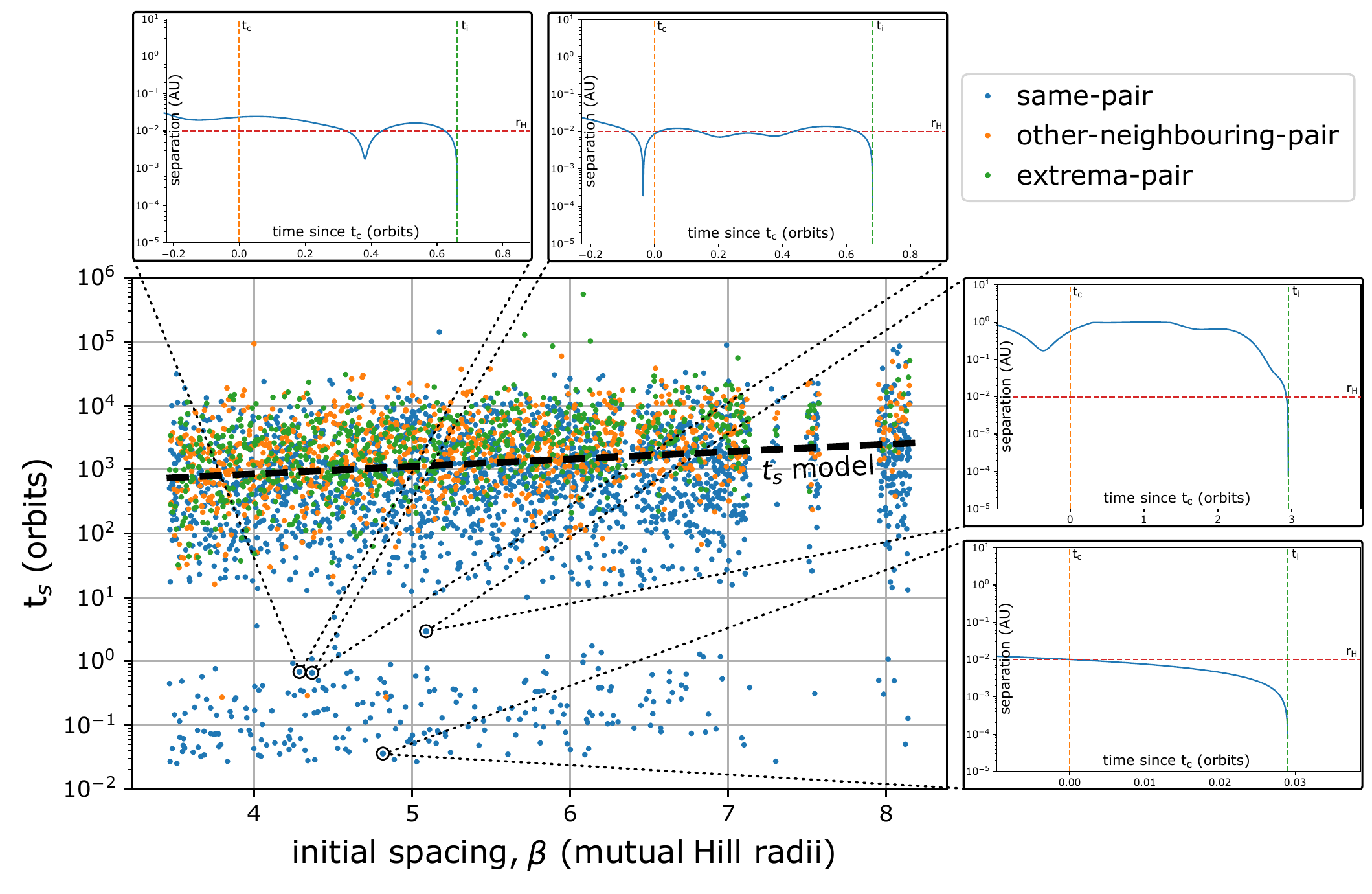}
    \caption{Post-crossing survival time of systems initially at $1$ AU against $\beta$. Blue dots indicate the same pair both crossed orbits and collided; orange indicates the pair that collided was not the pair that crossed; green indicates a collision between the inner and outer planets. The $t_s$ model (\rev{bold} dashed black) is fitted to all data points \rev{with a survival time greater than two orbits}. The insets show the planet separation for the marked systems between crossing time\rev{, $t_c$,} (dashed orange) and collision time\rev{, $t_i$,} (dashed green). Additionally, the Hill radius at $1$ AU is shown (dashed red).}
    \label{fig: survival time against beta with satellite images}
\end{figure*}

\begin{figure*}
    \centering
    \includegraphics[width=1.0\textwidth]{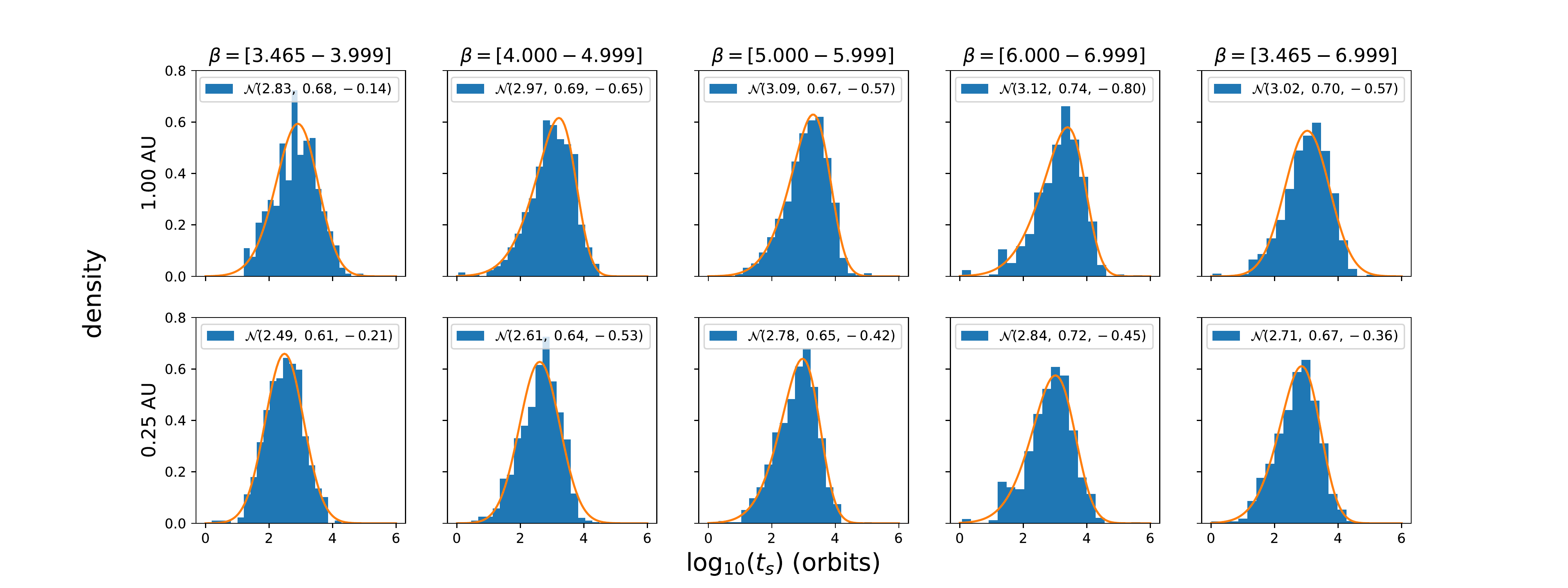}
    \caption{Normalised histograms of \rev{post-crossing survival time,} log($t_s$)\rev{, for different regions of initial spacing, $\beta$}. The top row of plots is for systems initially at $1$ AU while the bottom one is at $0.25$ AU. Log-skew-normal probability density functions, shown in orange, are fitted to the data through a maximum likelihood estimator. The mean $\mu$, standard deviation $\sigma$, and the skew $\zeta$ are included for each distribution as $\mathcal{N}(\mu, \sigma, \zeta)$. Systems that did not experience a crossing were excluded from these distributions.}
    \label{fig: survival distributions}
\end{figure*}

\section{Standard integration suite}
\label{sec: standard suite results}
This section contains results from the standard integration suite described in section \ref{sec: standard suite}. \rev{Additionally, the results of the perturbed integration suite, described in section \ref{sec: perturbed suite}, are analysed in subsection \ref{sec: sensitivity to initial conditions}.}

\subsection{Timescale to planet-planet collision}
\label{sec: time scale to planet planet collision}

 The crossing and impact times for the standard suite are plotted in Fig.~\ref{fig: orbital crossing at 1AU}. Inspection of the crossing time with respect to the initial orbital spacing shows the clear upwards trend present in other works \citep{Smith2009,Obertas2017,Hussain2020,Tamayo2020,Lissauer2021}. We also capture the large scale variations about the trend which for the most part are a result of mean motion resonances as discussed in \citet{Obertas2017}. Additionally, we replicate the finding of \citet{Lissauer2021} in the discovery of a highly stable configuration around $\beta=5.74$ which they attribute to the distance of this configuration from any strong resonances.

Throughout this work we used a linear logarithmic fit of the form
\begin{equation}
    \textrm{log}\rev{_{10}} \left(t \right) = b\rev{'} \beta' + c\rev{'}
    \label{eq: model}
\end{equation}
in several places where $\beta' = \beta-2\sqrt{3}$ and is used to reduce the dependency of \rev{the y-intercept} upon the slope. We fit this model to three data sets such that $t=$ $t_c$, $t_i$ or $t_s$ and state explicitly which at the time of use. \rev{Unless otherwise stated, w}e only include data points in the region $\beta = [3.465, 6.3]$ in the fits to avoid biasing the results due to systems that did not experience an orbital crossing within the maximum simulation time. For $t=t_c$, over this region, we find that  $b\rev{'} = 1.352$ and $c\rev{'} = 2.067$ which is in strong agreement with \citet{Lissauer2021} and confirms the functionality of the TES tool. For impact times $t_i$ we find $b\rev{'} = 1.192$ and $c\rev{'} = 2.42$.

Figure \ref{fig: orbital crossing at 1AU} highlights that the post-crossing survival time is very small compared to the crossing time for the majority of systems observed. The log scale of the plot and the relatively small magnitude of $t_s$ means the bulk of the impact time data points are hidden in this figure. The only exception is in the region of small $\beta$ where the ratio $\nicefrac{t_i}{t_c}$ is large due to the relatively small size of $t_c$.

Finally, it can be seen that for a small subset of integrations collisions can occur before an orbital crossing has taken place. A cumulative sum showing the numbers of occurrences is shown in Fig.~\ref{fig: cumsum ti before tc} where we believe that the increase between systems at $1~$AU and $0.25~$AU is not dependent purely on the physical cross-sectional area of planets but rather the enhanced cross-sectional area due to gravitational focusing \citep{Safronov1972}.
It is likely that a symplectic integrator, configured to use the standard step size of $\nicefrac{1}{20}$~th of the smallest dynamical period, would miss these collisions. However, given the small number of occurrences relative to the number of integrations typically performed in stability studies, it is unlikely that these missed collisions will have biased the data-sets in any statistically meaningful way. 

\begin{figure}
   \includegraphics[width=0.49\textwidth]{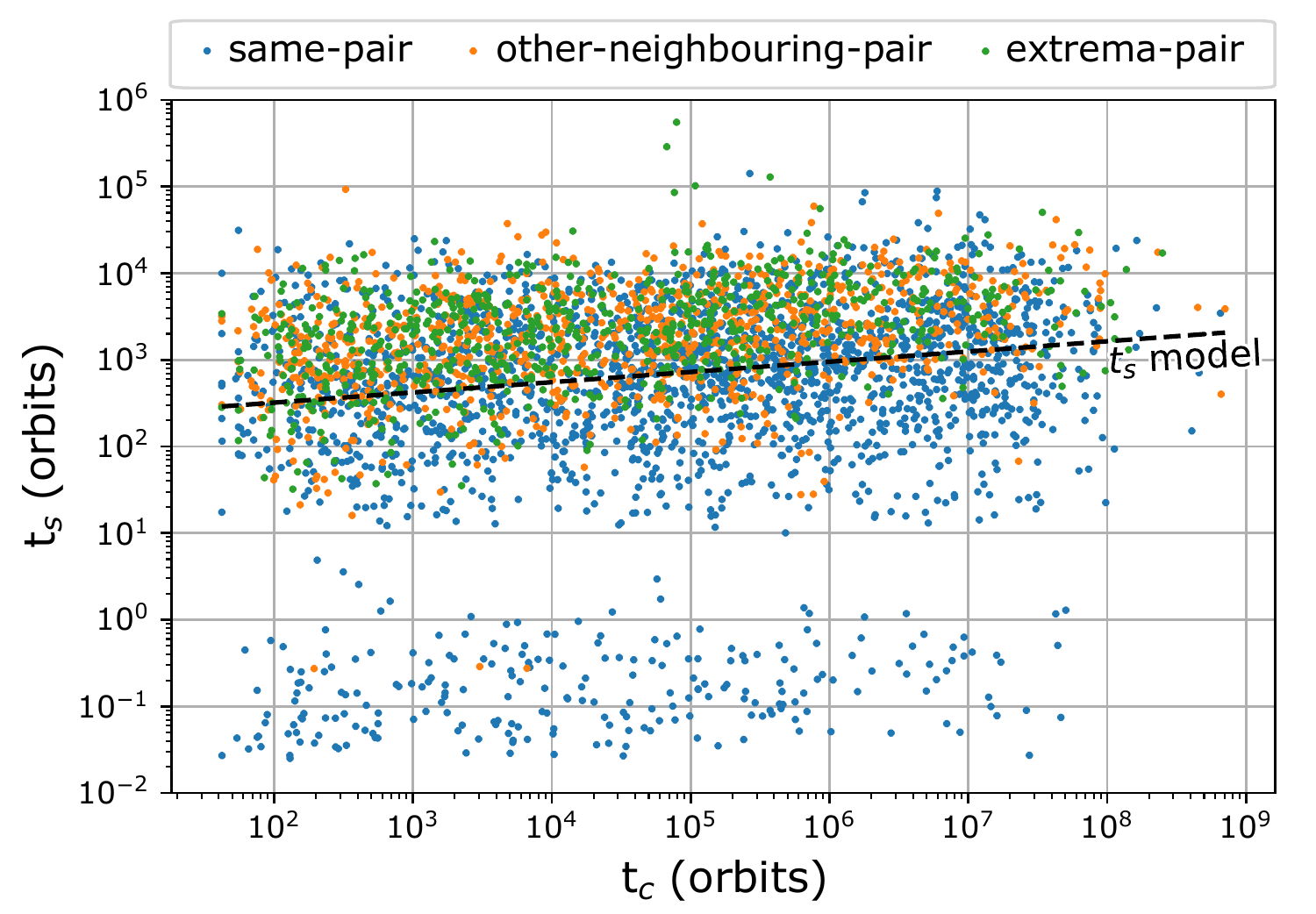}
\caption{Post-crossing survival time\rev{, $t_s$, against orbital crossing time, $t_c$, for} systems initially at $1$ AU. Blue dots indicate the same pair both crossed orbits and collided; orange indicates the pair that collided was not the pair that crossed; green indicates a collision between the inner and outer planets. The $t_s$ model (dashed black) is fitted to all data points \rev{with a survival time greater than two orbits}.}
\label{fig: survival times}
\end{figure}

\begin{table}
	\centering
	\caption{\rev{Fitted model coefficients for $t_s$ against $\beta$ and $t_c$. Plotted models are fitted to the long-lived population, long, only but fitted models for the full dataset, all, are included as well. $PCC$ is the Pearson correlation coefficient. $\sigma$ is the standard deviation of the dataset from the fitted model.}}
	\label{tab: model coefficients}
	\begin{tabular}{llcccccc} 
		\hline
		 $t_s$ model & dataset & $b$ & $c$ & $b'$ & $c'$ & $PCC$ & $\sigma$ \\
         \hline
        \multirow{2}{*}{Figure~\ref{fig: survival time against beta with satellite images}} 
        & long & $-$ & $-$ & $0.111$ & $2.84$ & $0.197$ & $0.680$    \\
        & all & $-$ & $-$ & $0.165$ & $2.496$ & $0.176$ & $1.13$    \\
        \multirow{2}{*}{Figure~\ref{fig: survival times}} 
         & long & $0.0781$ & $2.693$ & $-$ & $-$ & $0.183$ & $0.682$    \\
         & all & $0.118$ & $2.27$ & $-$ & $-$ & $0.167$ & $1.13$    \\
        \hline
	\end{tabular}
\end{table}

Figure~\ref{fig: survival time against beta with satellite images} shows the post-crossing survival time for all systems within the standard suite against $\beta$; Figure~\ref{fig: survival times} is identical but plotted against $t_c$. We find two main populations of post-crossing survival times present: those surviving for less than two orbits, and those surviving for more than ten orbits with very few outliers in between. Within the long surviving population, it can be seen that there is a clear increase in the post-crossing survival time of systems with respect to both $\beta$ and $t_c$. 
\rev{We fit models of the form of Eq.~\eqref{eq: model} to both the long-lived population and the population in its entirety, we call these datasets \emph{long} and \emph{all}, respectively. The model coefficients $b'$ and $c'$ can be found in the top two rows of Table~\ref{tab: model coefficients}. Similarly, we also fit linear models to the two datasets present in Fig.~\ref{fig: survival times} for $\mathrm{log}_{10}(t_s)$ against $\mathrm{log}_{10}(t_c)$. The model coefficients $b$ and $c$ can be found in the bottom two rows of Table~\ref{tab: model coefficients}. In all cases, we calculate the Pearson correlation coefficient (PCC) and also calculate the standard deviation, $\sigma$, of the data minus the fitted model, e.g. $\sigma(\mathrm{log}_{10}(t_s)-(b'\beta'+c'))$.
}
Clearly, there is a tendency for systems to persist for longer after an orbital crossing when the initial mutual spacing between them is greater, with a difference of a factor of three in median post-crossing survival time over the entire beta range. However, even given this increase, the post-crossing survival time for systems simulated did not ever exceed one million orbits. Given that this represents roughly one ten-thousandth of the main sequence lifetime of \rev{solar-mass} stars it is possible, although very unlikely, that we could observe a \rev{compact} exoplanet system that has undergone an orbital crossing but has not yet experienced a collision between planets, even if it were a truly co-planar system. 

In the case of the short\rev{-}lived population, there is a further subdivision of different behaviours: those systems that experience a collision almost immediately following a crossing, e.g. those bodies whereby $t_s$ < $10^{-1}$, and those which persist for longer than this but less than a couple of orbits. In the former case, we have observed that the trajectories of two planets about the star simply cross, leading to straightforward collisions, and also triggering an orbital crossing in the process. However, in the latter case, we find that the trajectories of the planets about the star are such that a very close encounter occurs which causes the two planets to become temporarily gravitationally captured. These two planets then remain within approximately a Hill radius of one another before finally experiencing a fatal collision a fraction of an orbit later. These behaviours are shown in the satellite images in Fig.~\ref{fig: survival time against beta with satellite images}. It can be seen here that temporary gravitational capture is not the cause of collision in the case of the outliers with a post-crossing survival time between two and ten orbits.

\begin{figure}
    \includegraphics[width=0.495\textwidth]{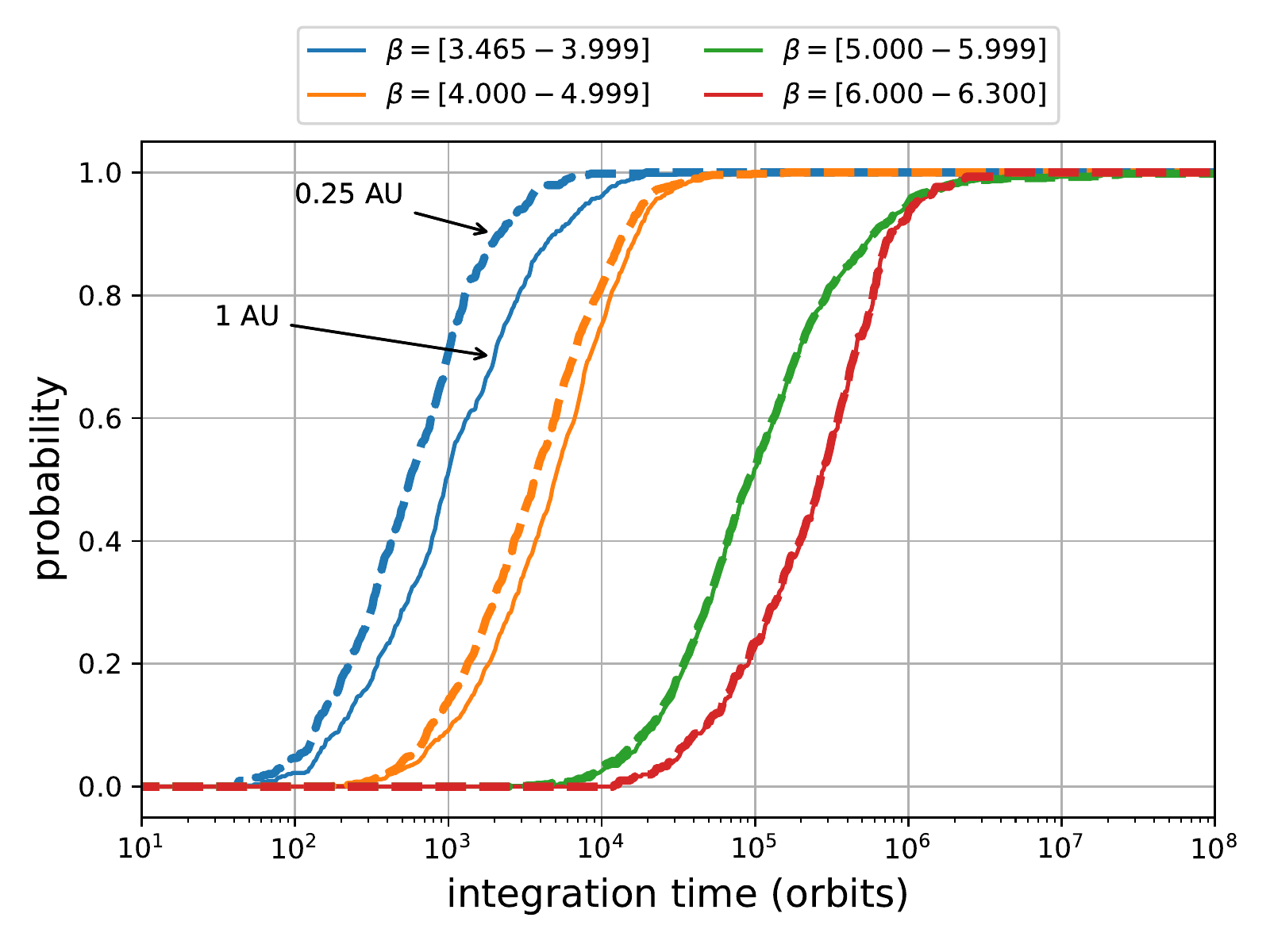}
\caption{Probability of having experienced a collision over time for various regions of \rev{initial spacing,} $\beta$. \rev{The probability is calculated as the cumulative fraction of systems that have experienced collisions over the total number of systems.} Solid lines show the probabilities  for systems initially at $1$ AU while the dashed lines are initially at $0.25$ AU.}
\label{fig: collision probabilities}
\end{figure}

To give consideration to whether these results generalise to other systems of planets we have calculated $t_i$ and $t_s$ for a system with the inner planet initially placed at $0.25$ AU.
\rev{This is equivalent to artificially inflating the radius of all planets in systems at $1$~AU by a factor of four. When thought of this way, this is akin to placing planets} with a radius approximately the same size as Neptune at $1$ AU; $t_c$ is invariant to the initial location of the inner planet. The probability\rev{, calculated as the cumulative fraction of systems that have experienced collisions over the total number of systems,} of collision over time for both settings are shown in Fig.\ref{fig: collision probabilities}. The separation between dashed and solid lines indicates that a given collision probability is reached sooner in systems composed of planets with a larger radius. The difference in time remains about constant over all values of $\beta$, even if the log scale suggests otherwise. 

Figure \ref{fig: survival distributions} contains \rev{normalised histograms} of $t_s$ within different regions of $\beta$ for systems with the inner planet initially at $1$ AU and $0.25$ AU. We find that the distribution of post-crossing survival times is log-skew-normal distributed across all systems; we confirmed this using a Kolmogorov-Smirnov test with a precision parameter of $\alpha=0.005$. The skew-normal distribution is a generalisation of the normal distribution that allows the class to be extended to include distributions with non-zero skewness through the addition of a shape parameter \citep{Azzalini1999}. Log-skew-normal probability density functions, shown in orange, are fitted to the data through a maximum likelihood estimator. We calculated the mean $\mu$, standard deviation $\sigma$, and the skew $\zeta$ for each distribution; we use the Fisher-Pearson coefficient of skewness throughout. We find that $\mu$ increases with increasing $\beta$ range, and also find the same pattern for $\sigma$ in all but one case.
In all cases, $\zeta$ is negative indicating a skew towards shorter post-crossing survival times as compared to a normal distribution. 
\rev{This means that there is a preference for systems to collide sooner rather than later after an orbital crossing as compared to the most frequent survival times. There is a slow build up in the number of systems experiencing collisions over time after an orbital crossing but a much sharper cut-off after the peak density of collisions. This highlights the difficulty for systems to persist for long timescales after an orbital crossing in the co-planar case.}
Systems with a shorter mean post-crossing survival time show a skew of a smaller magnitude than those with a longer survival time, e.g. at $1$ AU $\zeta=-0.14$ for $\beta < 4.0$ whereas for $\beta >= 4.0$ the smallest, in magnitude, value observed is $\zeta=-0.57$.
We find that the distributions of post-crossing survival times at $0.25$ AU are less skewed than those at $1$ AU, indicating that the survival times of systems in this case are closer to a log-normal distribution.

\begin{table*}
	\centering
	\caption{Comparison of \emph{crossing times} of systems using identical values of \rev{initial spacing, $\beta$, in mutual Hill radii} for the standard and perturbed initial longitudes.}
	\label{tab:crossing table}
	\begin{tabular}{lcccccc} 
		\hline
		Interval: & [$3.465$, $3.999$] & [$4.0$, $4.999$] &  [$5.0$, $5.999$] & [$6.0$, $6.33$] &  \hspace{0.3cm} & [$3.46\rev{5}$, $6.33$]\\
        \hline
        number of runs in the range & $535$  & $1000$  & $1000$  & $331$  && $2866$  \\
        $< \textrm{log}_{t_c}\textrm{(standard)} - \textrm{log}_{t_c}\textrm{(perturbed)}>$ & $0.006$  & $-0.001$  & $-0.011$  & $-0.014$  && $-0.004$  \\
        $< |\textrm{log}_{t_c}\textrm{(standard)} - \textrm{log}_{t_c}\textrm{(perturbed)} |>$ & $0.039$  & $0.182$  & $0.306$  & $0.356$  && $0.219$  \\
        $t_c$(perturbed) < $0.5 t_c$(standard) & $7$ ($1.31\%$)  & $92$ ($9.20\%$)  & $200$ ($20.00\%$)  & $75$ ($22.66\%$)  && $374$ ($13.05\%$)  \\
        $0.5 t_c$(standard) < $t_c$(perturbed) < $2 t_c$(standard) & $524$ ($97.94\%$)  & $812$ ($81.20\%$)  & $580$ ($58.00\%$)  & $173$ ($52.27\%$)  && $2089$ ($72.89\%$)  \\
        $t_c$(standard) < $0.5 t_c$(perturbed) & $4$ ($0.75\%$)  & $96$ ($9.60\%$)  & $220$ ($22.00\%$)  & $83$ ($25.08\%$)  && $403$ ($14.06\%$)  \\
        within 10\% of standard systems & $398$ ($74.39\%$)  & $217$ ($21.70\%$)  & $100$ ($10.00\%$)  & $27$ ($8.16\%$)  && $742$ ($25.89\%$)  \\
        within 1\% of standard systems & $333$ ($62.24\%$)  & $68$ ($6.80\%$)  & $10$ ($1.00\%$)  & $7$ ($2.11\%$)  && $418$ ($14.58\%$)  \\
        \hline 
	\end{tabular}
\end{table*}

\begin{table*}
	\centering
	\caption{Comparison of \emph{collision times} of systems using identical values of \rev{initial spacing, $\beta$, in mutual Hill radii} for the standard and perturbed initial longitudes both with the innermost planet initially at \emph{$1$ AU}.}
	\label{tab:collision table}
	\begin{tabular}{lcccccc} 
		\hline
	Interval: & [$3.465$, $3.999$] & [$4.0$, $4.999$] &  [$5.0$, $5.999$] & [$6.0$, $6.33$] &  \hspace{0.3cm} & [$3.46\rev{5}$, $6.33$]\\
    \hline
    number of runs in the range & $535$  & $1000$  & $1000$  & $331$  && $2866$  \\
    $<\textrm{log}_{t_i}\textrm{(standard)} - \textrm{log}_{t_i}\textrm{(perturbed)}>$ & $0.015$  & $-0.012$  & $-0.010$  & $-0.012$  && $-0.006$  \\
    $<|\textrm{log}_{t_i}\textrm{(standard)} - \textrm{log}_{t_i}\textrm{(perturbed)} |>$ & $0.429$  & $0.302$  & $0.294$  & $0.349$  && $0.328$  \\
    $t_i$(perturbed) < $0.5 t_i$(standard) & $145$ ($27.10\%$)  & $189$ ($18.90\%$)  & $185$ ($18.50\%$)  & $76$ ($22.96\%$)  && $595$ ($20.76\%$)  \\
    $0.5 t_i$(standard) < $t_i$(perturbed) < $2 t_i$(standard) & $260$ ($48.60\%$)  & $614$ ($61.40\%$)  & $598$ ($59.80\%$)  & $175$ ($52.87\%$)  && $1647$ ($57.47\%$)  \\
    $t_i$(standard) < $0.5 t_i$(perturbed) & $130$ ($24.30\%$)  & $197$ ($19.70\%$)  & $217$ ($21.70\%$)  & $80$ ($24.17\%$)  && $624$ ($21.77\%$)  \\
    within 10\% of standard systems & $108$ ($20.19\%$)  & $110$ ($11.00\%$)  & $99$ ($9.90\%$)  & $25$ ($7.55\%$)  && $342$ ($11.93\%$)  \\
    within 1\% of standard system & $79$ ($14.77\%$)  & $17$ ($1.70\%$)  & $8$ ($0.80\%$)  & $5$ ($1.51\%$)  && $109$ ($3.80\%$)  \\
    \hline
	\end{tabular}
\end{table*}

\begin{table*}
	\centering
	\caption{Comparison of \emph{collision times} of systems using identical values of \rev{initial spacing, $\beta$, in mutual Hill radii} for the standard and perturbed initial longitudes both with the innermost planet initially at \emph{$0.25$ AU}.}
	\label{tab: collision table 0.25 au}
	\begin{tabular}{lcccccc} 
		\hline
		Interval: & [$3.465$, $3.999$] & [$4.0$, $4.999$] &  [$5.0$, $5.999$] & [$6.0$, $6.33$] &  \hspace{0.3cm} & [$3.46\rev{5}$, $6.33$]\\
\hline
number of runs in the range & $535$  & $1000$  & $1000$  & $331$  && $2866$  \\
$<\textrm{log}_{t_i}\textrm{(standard)} - \textrm{log}_{t_i}\textrm{(perturbed)}>$ & $-0.005$  & $-0.010$  & $-0.010$  & $-0.011$  && $-0.009$  \\
$<|\textrm{log}_{t_i}\textrm{(standard)} - \textrm{log}_{t_i}\textrm{(perturbed)} |>$ & $0.297$  & $0.243$  & $0.301$  & $0.353$  && $0.286$  \\
$t_i$(perturbed) < $0.5 t_i$(standard) & $98$ ($18.32\%$)  & $141$ ($14.10\%$)  & $187$ ($18.70\%$)  & $75$ ($22.66\%$)  && $501$ ($17.48\%$)  \\
$0.5 t_i$(standard) < $t_i$(perturbed) < $2 t_i$(standard) & $335$ ($62.62\%$)  & $701$ ($70.10\%$)  & $592$ ($59.20\%$)  & $176$ ($53.17\%$)  && $1804$ ($62.94\%$)  \\
$t_i$(standard) < $0.5 t_i$(perturbed) & $102$ ($19.07\%$)  & $158$ ($15.80\%$)  & $221$ ($22.10\%$)  & $80$ ($24.17\%$)  && $561$ ($19.57\%$)  \\
within 10\% of standard systems & $142$ ($26.54\%$)  & $130$ ($13.00\%$)  & $99$ ($9.90\%$)  & $25$ ($7.55\%$)  && $396$ ($13.82\%$)  \\
within 1\% of standard system & $108$ ($20.19\%$)  & $19$ ($1.90\%$)  & $10$ ($1.00\%$)  & $7$ ($2.11\%$)  && $144$ ($5.02\%$)  \\
\hline
	\end{tabular}
\end{table*}

\subsection{Sensitivity to initial conditions}
\label{sec: sensitivity to initial conditions}
To examine the sensitivity to initial conditions of the results of our simulations we use our perturbed suite of integrations described in section \ref{sec: perturbed suite}. The crossing and collision times of each integration between the standard suite and the perturbed suite are compared to determine the effect of the perturbation. Table \ref{tab:crossing table} contains the results of that comparison for the \emph{time of orbital crossing}. Tables \ref{tab:collision table} and \ref{tab: collision table 0.25 au} contain the same comparison for but for the \emph{impact time} of systems at $1$ AU and $0.25$ AU, respectively.

In general, the comparison between crossing times in Table \ref{tab:crossing table} aligns closely with \cite{Lissauer2021}. Percentages between the two studies rarely differ by more than a few points despite the different integration tools used: TES and MERCURY. One notable difference between the two studies is in the \rev{initially wider spaced} systems. In the regions $\beta = [5.0, 5.999]$ and $\beta =  [6.0, 6.33]$ we find roughly double the number of initial orbital spacings where the standard and perturbed suite integrations experience orbital crossing times within $10\%$ of one another.
Given the precise orbital evolution required in order for standard and perturbed suite systems to experience a crossing at the same time it is unlikely that numerical error would ever cause an increase in this statistic. We therefore take this as an  indication that TES has maintained a higher precision than the  symplectic Wisdom-Holman \citep{Wisdom1991} scheme within MERCURY. To further validate TES in this setting we have also repeated the standard suite integrations with \rev{IAS15} from the REBOUND package for comparison. We find very good agreement in results between the two routines. Detailed results from this experiment are included in Appendix \ref{appendix: integrator comparison}.

The right-most summary column for the full range of $\beta = [3.465, 6.33]$ in Table \ref{tab:collision table} shows there is a marked decrease in the number of collisions occurring within a factor of two, and within ten and one percent of one another; as compared to the orbital crossing times in Table \ref{tab:crossing table}. The largest reduction is seen in the within\rev{-}a\rev{-}factor\rev{-}of\rev{-}two row where a reduction of over $15$ \rev{percentage points} highlights the sensitivity to close approaches in this setting.
The majority of this difference in collision times is seen in the \rev{initially closely spaced} systems where a reduction of \rev{almost} $50$ \rev{percentage points} can be seen for integrations finishing within $10$\% of one another. However, once the crossing times exceed approximately $1 \times 10^4$ orbits at $\beta = 5$ the effect of the perturbation disappears and values between crossing and collision times for the two data sets converge. 

\begin{figure}
    \centering
    \includegraphics[width=0.475\textwidth]{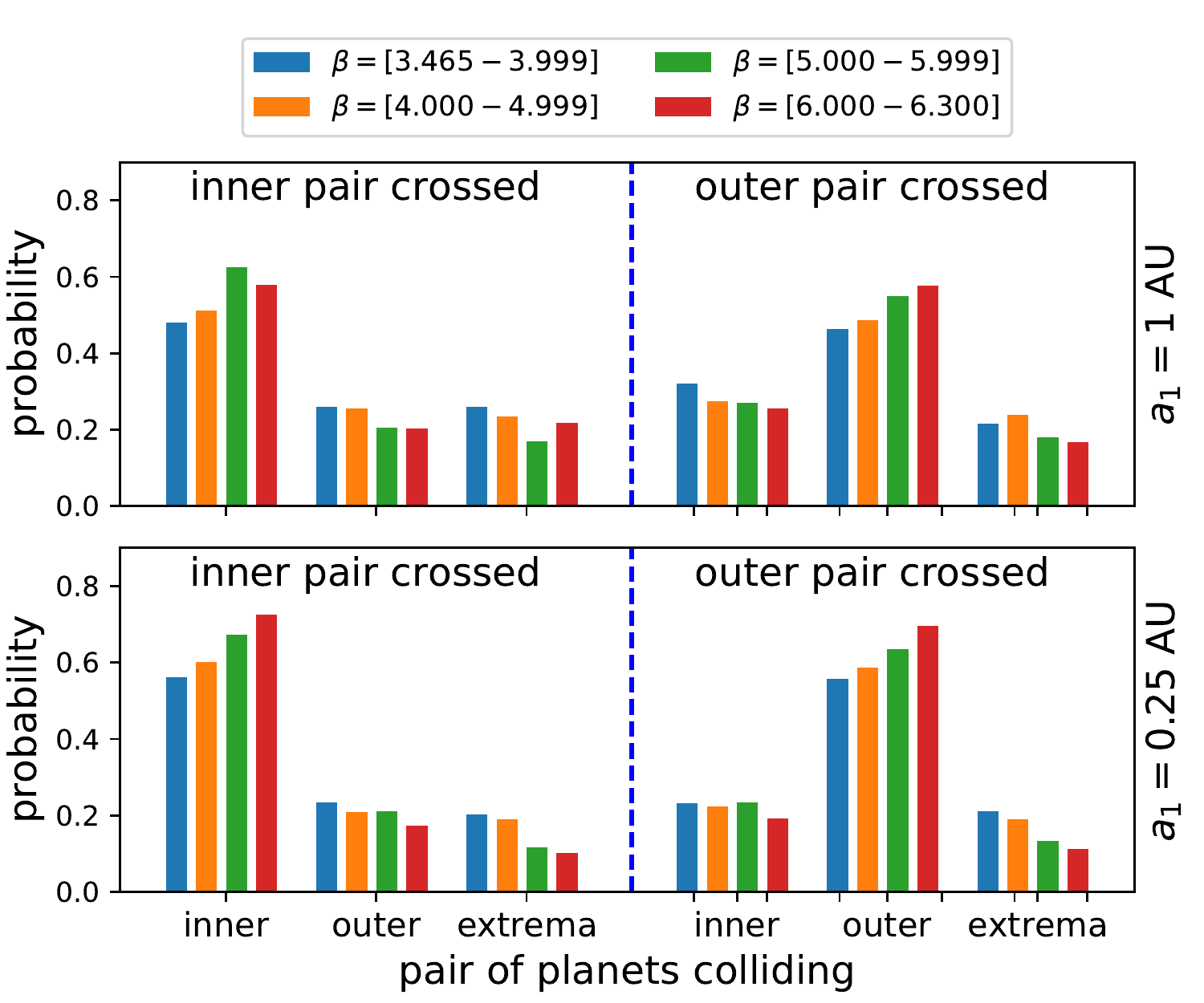}
    \caption{Probability of collision per pair of planets broken down by the pair of orbits that initially crossed and \rev{initial spacing,} $\beta$, range. \rev{Probability is calculated as the fraction of collisions between a given pair of planets over the total number of collisions.}The top pane\rev{l} is for systems initially at $1$ AU while the bottom pane\rev{l} is initially at $0.25$ AU. Inner and outer refer to the innermost and outermost pairs of planets, respectively. Extrema refers to the pair comprising the innermost and outermost planets.}
    \label{fig: proability collision from crossing 1 AU}
\end{figure}

\begin{figure}
    \centering
    \includegraphics[width=0.475\textwidth]{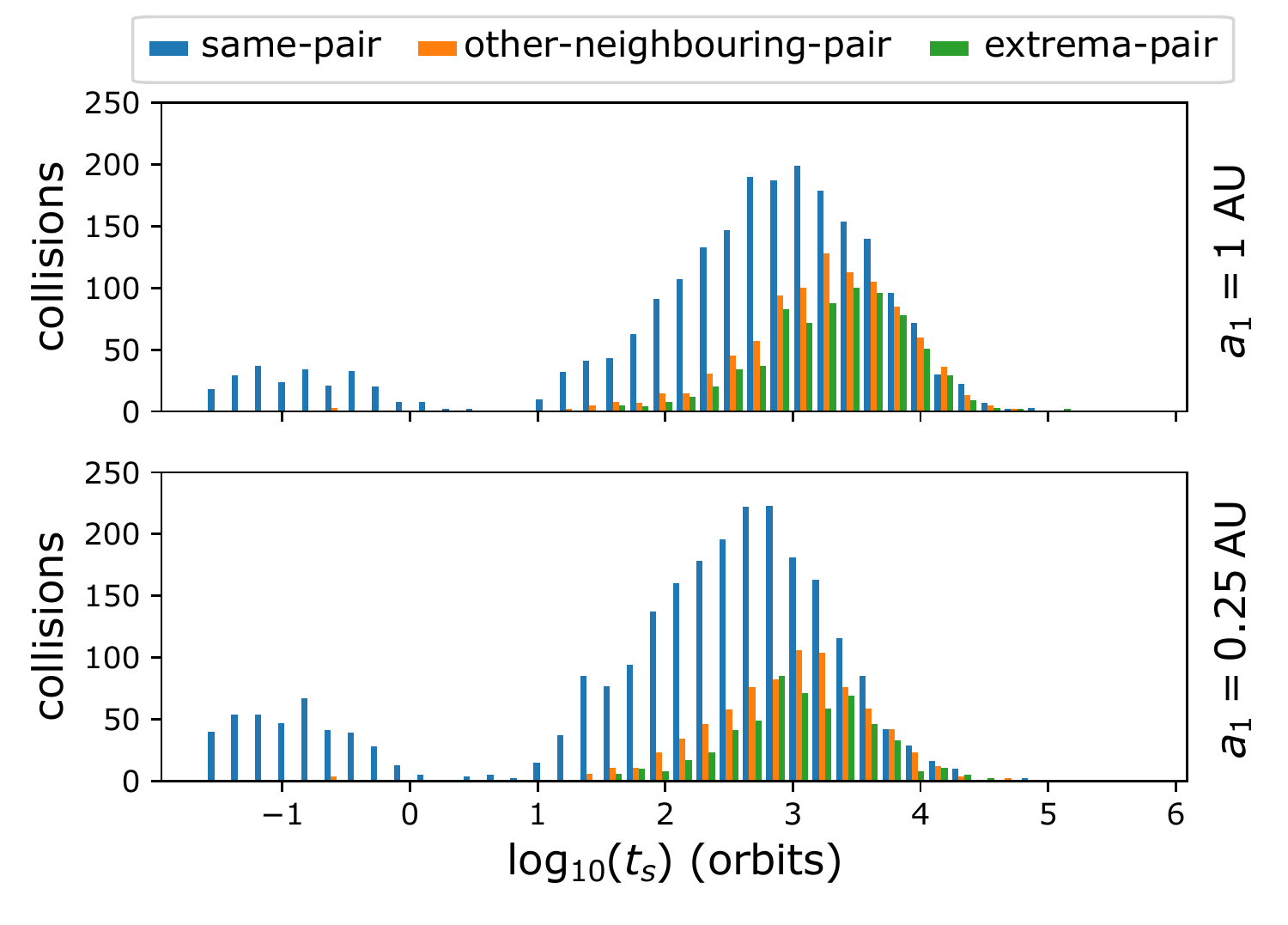}
    \caption{\rev{Post-crossing survival time} distribution of collisions between different  pairs of planets. Blue bars indicate the same pair both crossed orbits first and collided; orange indicates the pair that collided was  the other neighbouring pair; green indicates a collision between the inner and outer planets. The top pane\rev{l} is initially at $1$ AU while the bottom pane\rev{l} is initially at $0.25$ AU.}
    \label{fig: pairs involved in collision distribution}
\end{figure}

\subsection{Which planets collide?}
\label{sec: which planets collide}
We find a slight discrepancy between the prevalence of orbital crossings \rev{in our standard integration suite}, \rev{with the innermost pair triggering $48\%$ of crossings compared to $52\%$ for the outermost pair.} These percentages were calculated using $n=4,835$ integrations and the expected stochastic variation, about the mean, i.e. $50\%$, is therefore approximately $\rev{0.72}\%$ \citep{Dobrovolskis2007}.

In the following, we designate the specific pair of planets that collide as the \emph{collision pair}, and analogously we refer to the pair of planets that experienced an orbital crossing as the \emph{crossing pair}. We find that across all values of $\beta$ a collision between two planets is almost twice as likely if the same two planets were also involved in the orbital crossing. Figure \ref{fig: proability collision from crossing 1 AU} highlights clearly that this is the case with between $48$\% and $62$\% of collision events occurring between the crossing pair \rev{for systems at $1$~AU depending on the initial orbital spacing}. Moreover, these percentages appear to be invariant as to whether the inner or outer pair was involved in the orbital crossing. A clear trend can be seen with respect to $\beta$, where an increase in the initial orbital spacing between planets leads to an increased probability of collision between the crossing pair.

Figures \ref{fig: survival time against beta with satellite images} and \ref{fig: survival times} are coloured based on the collision and crossing pair for each system. As first crossings can only ever occur between neighbouring planets, it is possible to use only three colours for this: blue for \emph{same-pair systems} whereby the same pair was involved in both the first orbit crossing and the collision, orange for \emph{other-neighbouring-pair systems} to indicate that the colliding pair was the neighbouring pair that did not first cross, and green for \emph{extrema-pair systems} to indicate a collision between the inner and outer planets. Across the whole range of $\beta$ it can be seen that for systems with a $t_s$ below the $t_s$ model fit line collisions are predominantly between the first crossing pair. Figure \ref{fig: pairs involved in collision distribution} shows how these three combinations of events are distributed over time. Collisions that take place within a single orbit of orbit crossing are almost exclusively found in same-pair systems due either to simple immediate collisions or to the temporary gravitational bounding of planets discussed previously. Same-pair collisions are the most likely outcome for all systems at $1$ AU, shown in the top pane, until $t_s \approx 10^4$ orbits, followed by other-\rev{neighbouring-}pair systems, with extrema-pair systems being the least likely. However, after this period the probability of collision between any combination of planets becomes almost identical, indicating that the  mixing of planetary orbits after crossing is sufficient to overcome the increased probability of same-pair integrations due to the initial orbital configuration. Interestingly, the peak of other-\rev{neighbouring-}pair and extrema-pair systems do not align, instead  the former peaks first. This can be understood as the mixing process taking longer to cause the inner and outer planets orbits to overlap than to excite the middle planet enough to cross the orbits of both of its neighbours.
In the bottom pane, it can be seen that at $0.25$ AU the behaviour is similar; however, the number of collisions taking place within a single orbit roughly doubles. 

\section{Inclined Integration Suite}
\label{sec: inclined integration suite}

In the co-planar case, no systems survived for more than a million orbits after the first orbital crossing. However, \citet{Rice2018} observed a number of non co-planar systems that survived for their maximum simulation time of ten million orbits. Therefore, we now go on to examine the behaviour in the non co-planar case described by the inclined suite of initial conditions in section \ref{sec: inclined suite}. As a reminder, these initial conditions include fifteen initial inclinations ranging from an initial orbital height of $0.10 ~ r_H$ to $r_H$. 

\begin{figure}
    \centering
    \includegraphics[width=0.475\textwidth]{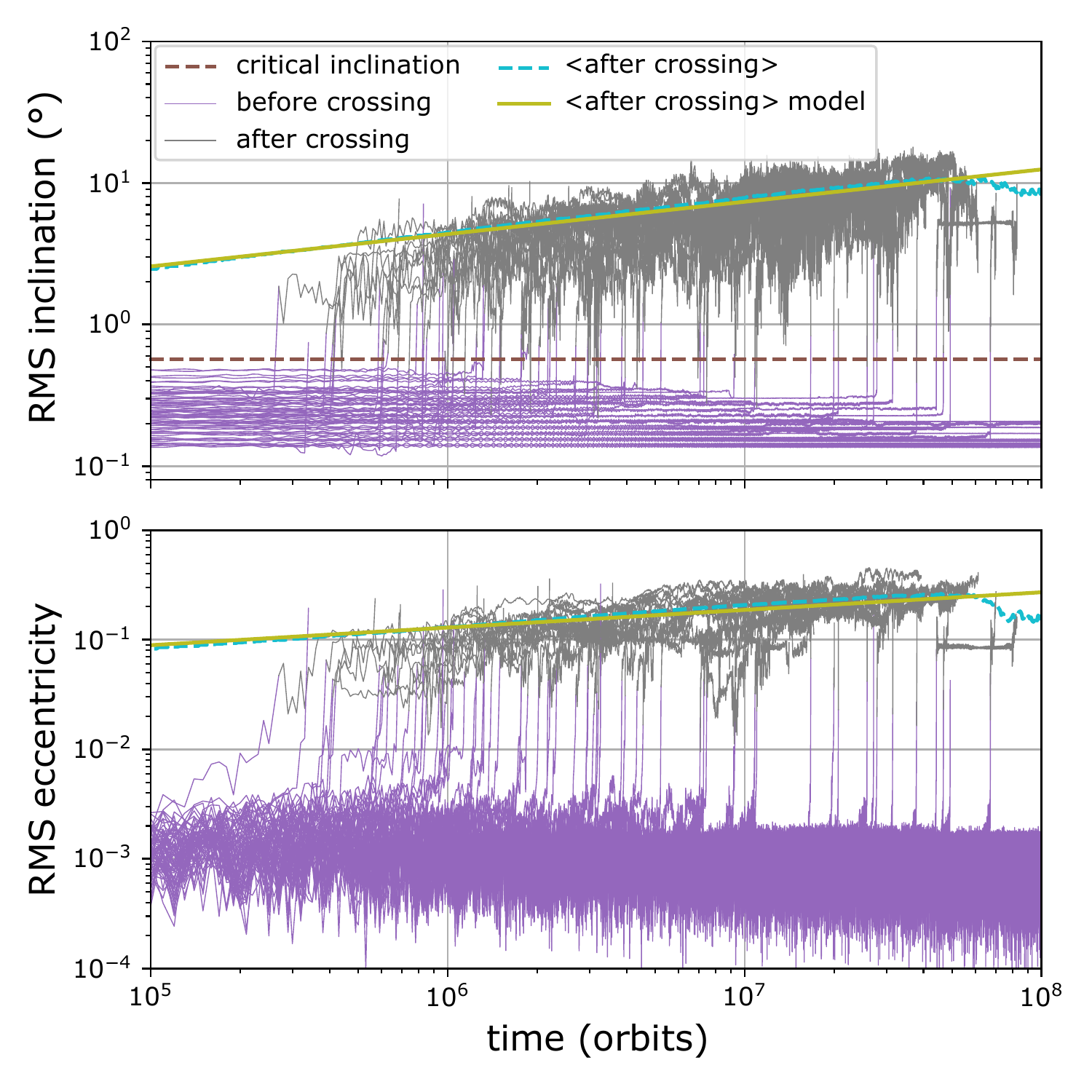}
    \caption{Inclination and eccentricity growth for individual systems from the inclined suite with $\beta = 5.98$. Only eighty configurations are included to aid clarity.
    Systems are shown in purple until they experience an orbital crossing and in grey thereafter. \rev{The RMS inclination and eccentricity values for all systems that have experienced an orbital crossing are shown (dashed blue). A linear model fitted to the mean of all systems that have experienced an orbital crossing is also shown (solid green).}}
    \label{fig: rms eccenticity inclination}
\end{figure}

\subsection{Dynamic heating}
\label{sec: dynamic heating}
The systems studied in the inclined integration suite begin with modest inclinations and no eccentricities, making them dynamically cold. Figure \ref{fig: rms eccenticity inclination} shows how the system heats up over time by plotting the root-mean-square (RMS) inclination and eccentricity over time. We calculate the mean over all runs that have experienced an orbital crossing, and fit a linear model to this mean \rev{which is shown as the solid green line}. Individual integrations  are shown in purple until they experience an orbital crossing and in grey thereafter. For clarity, in Fig.~\ref{fig: rms eccenticity inclination} results of individual integrations are only shown for eighty integrations in the inclined suite for $\beta = 5.98$.

\citet{Rice2018} found that, for four-planet Neptune-mass systems, there are two distinct growth modes of RMS eccentricity before and after an instability event: Eccentricity evolves rapidly to a quasi-equilibrium at a value of $~10^{-2}$ at which point encounters begin. After a period of mixing as a result of close approaches, systems transition into a new evolutionary phase during which eccentricity growth follows a power-law form approximately $\propto t^{\nicefrac{1}{6}}$. In the three-planet Earth-mass case, our systems reach a quasi-equilibrium value of $e \approx \,10^{-3}$ before a period of chaotic mixing and rapid growth, which finally settles into the new growth phase approximately $\propto t^{\nicefrac{1}{6}}$.

The RMS inclinations in Fig.~\ref{fig: rms eccenticity inclination}, on the other hand, while similar are different to the four-planet Neptune-mass case. We also observe that the inclination of the systems remains at roughly the initial value until the first encounter, at which point they are rapidly excited before entering a new growth mode. This rapid excitation is in keeping with the findings of \cite{Matsumoto2017}. These behaviours can be seen by the horizontal inclination lines in the population of systems before crossing and in the power-law growth in the population afterward. \citet{Rice2018} stated that the trend towards long-lived systems depends upon only the RMS inclination being greater than the averaged ratio of Hill radius to semi-major axis, this is called the critical inclination and is marked on this plot. We also find this to be the case across all systems within our inclined suite: any systems that have experienced orbital crossing and have their RMS inclination damped below this threshold rapidly experience a collision. The key difference in results in our simulations as compared to the four-planet, Neptune-mass case is that the power-law growth rate appears to be $\propto t^{\nicefrac{1}{4}}$ as opposed to $\propto t^{\nicefrac{1}{3}}$. We offer two possible explanations for this: 1) our data set could be biased due to the non-random initial conditions used; or, 2) there could be an underlying dependence between either the planetary mass or the number of planets within the system and the growth rate. Further investigation is needed to distinguish between these two possibilities.

\begin{figure}
    \centering
    \includegraphics[width=0.475\textwidth]{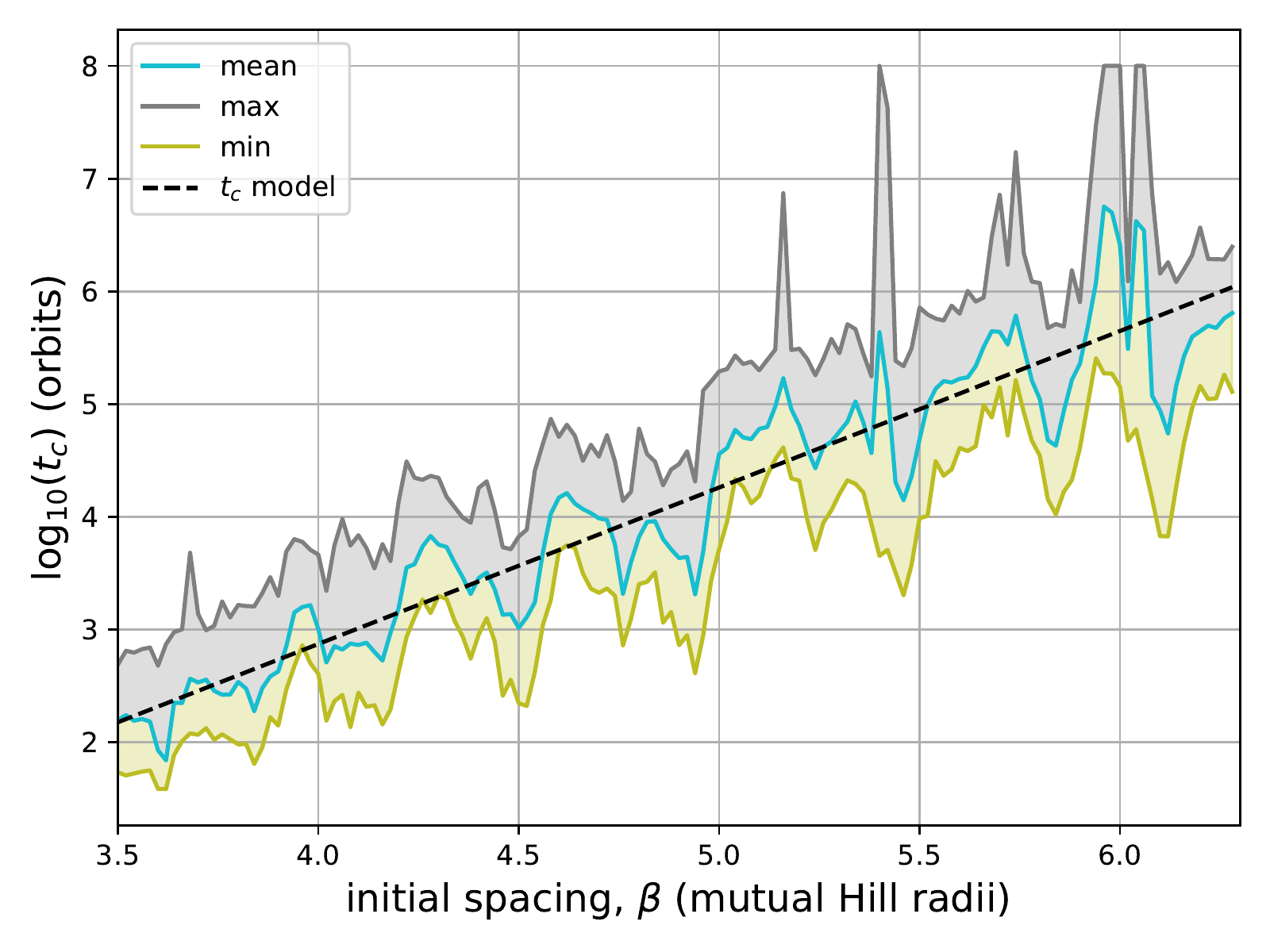}
    \caption{Time to orbital crossing against $\beta$ for the inclined integration suite. The minimum, maximum and mean values of the \rev{one hundred and twenty} integrations performed at each value of $\beta$ are shown. Additionally, the $t_c$ model is fitted to the mean values.}   
    \label{fig: tc inclined mean min max}
\end{figure}

\begin{figure}
    \centering
    \includegraphics[width=0.475\textwidth]{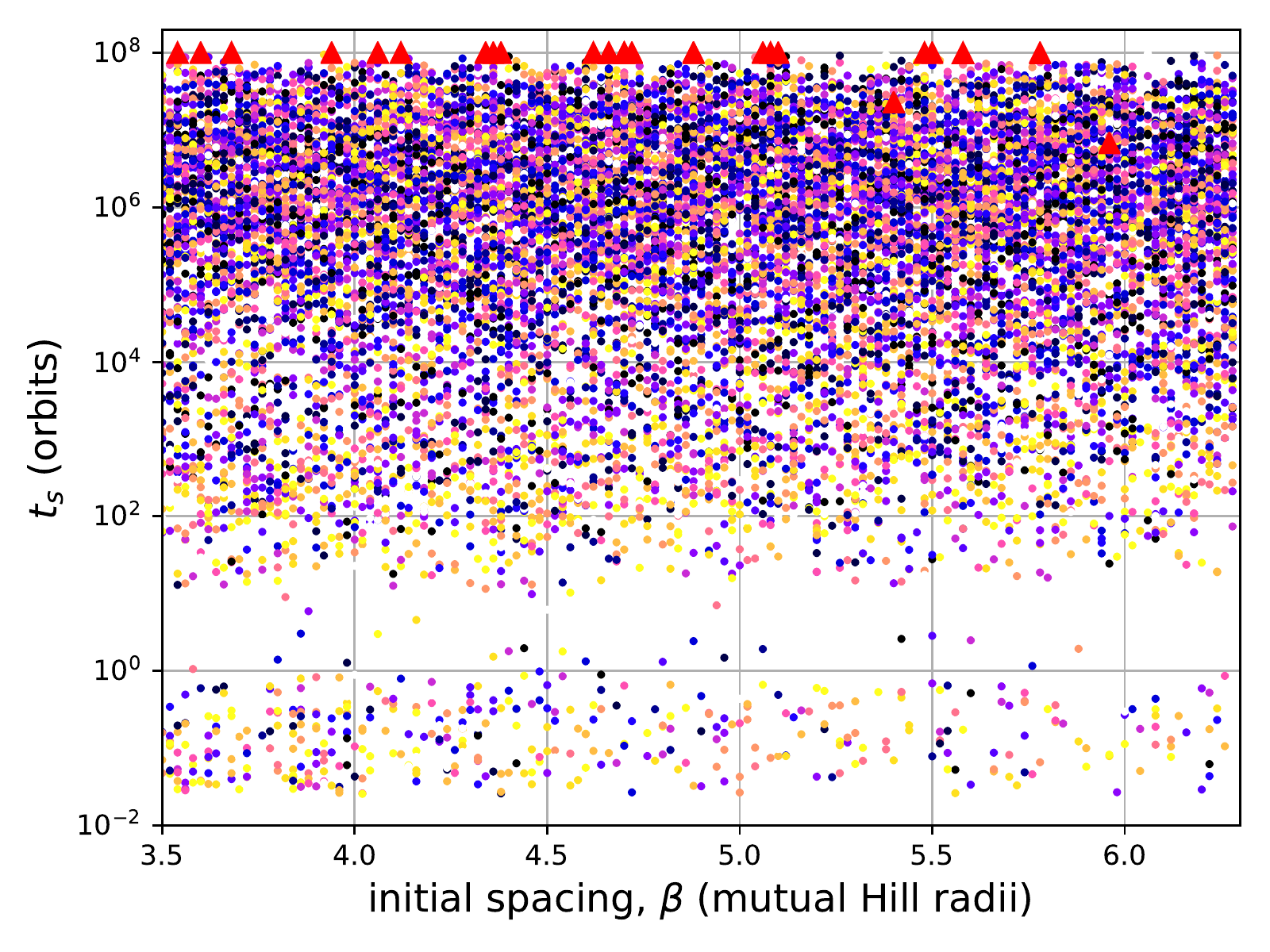}
    \caption{Post-crossing survival time of inclined integration suite with systems at $1$ AU. Colours \rev{of data points represent the initial inclinations, with darker colours representing higher inclinations}. The twenty-three systems that persisted for the full $10^8$ orbits are highlighted via a red triangle, independent of their initial inclination. Note that most of these surviving systems had their initial orbital crossing in far less than $10^8$ years, so they survived for almost $10^8$ years post-crossing before the simulation was terminated and appear as triangles at the top of the plot; the two exceptions, which survived for $< 3 \times 10^7$ years, both had initial orbital separations $\beta > 5.3$. }
    \label{fig: ts inclined}
\end{figure}

\begin{figure}
    \centering
    \includegraphics[width=0.475\textwidth]{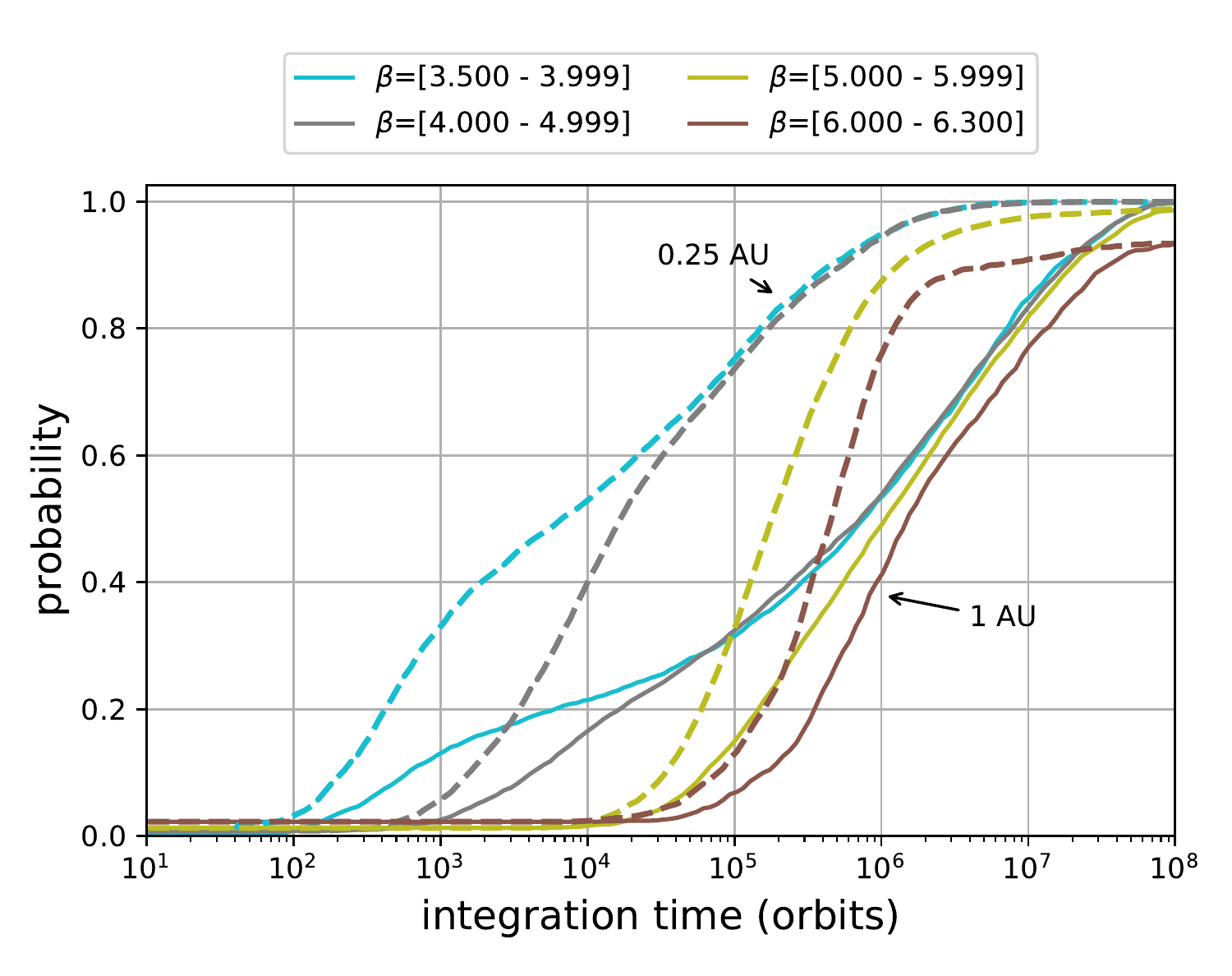}
    \caption{Probability of having experienced a collision over time for various regions of $\beta$ in the inclined integration suite. \rev{The probability is calculated as the cumulative fraction of systems that have experienced collisions over the total number of systems.} Solid lines show the probabilities  for systems initially at $1$ AU while the dashed lines are initially at $0.25$ AU.}
    \label{fig: ti inclined probability}
\end{figure}

\begin{figure}
    \centering
    \includegraphics[width=0.475\textwidth]{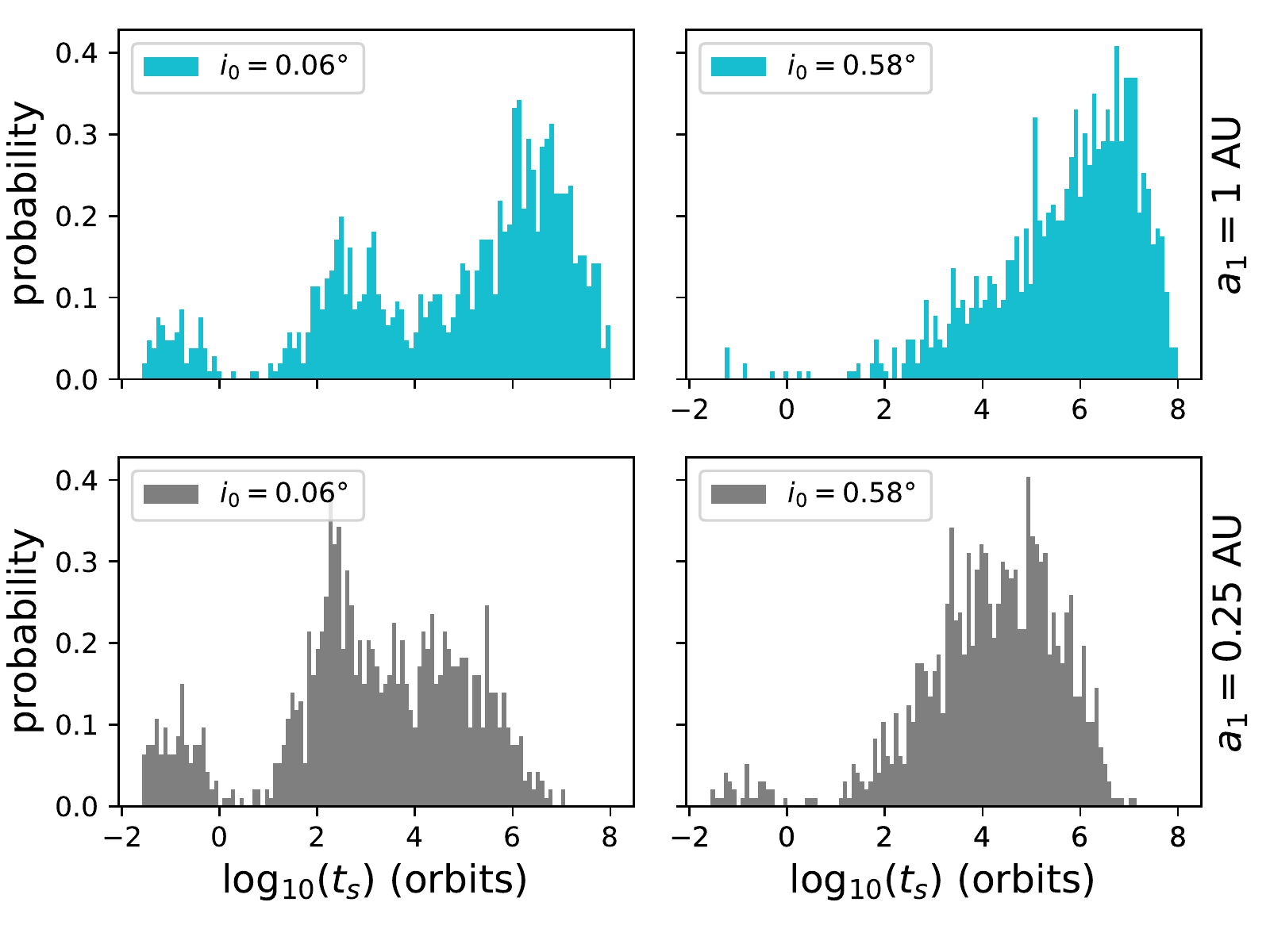}
    \caption{Distribution of post\rev{-crossing} survival times in the inclined integration suite for systems after a close encounter. Each plot contains data from $1120$ integrations across the entire inclined $\beta$ range where $\beta = 3.5-6.3$. The upper two plots, in cyan, are for systems initially at $1~$AU and the lower two plots, in grey, are for $0.25~$AU. The two leftmost plots contain data for systems with the minimum initial inclination, $i_0 = 0.06^\circ$, whereas the two rightmost plots contain data for systems with the maximum initial inclination, $i_0 = 0.58^\circ$. Two systems survived for the full simulation time after an orbital crossing in the low inclination case at $1~$AU whereas one survived in the high inclination case. No systems in the $0.25~$AU case survived for the full simulation duration after an orbital crossing in any of our integrations.}
    \label{fig: inclined survival probabilities four panel}
\end{figure}

\begin{figure}
    \centering
    \includegraphics[width=0.475\textwidth]{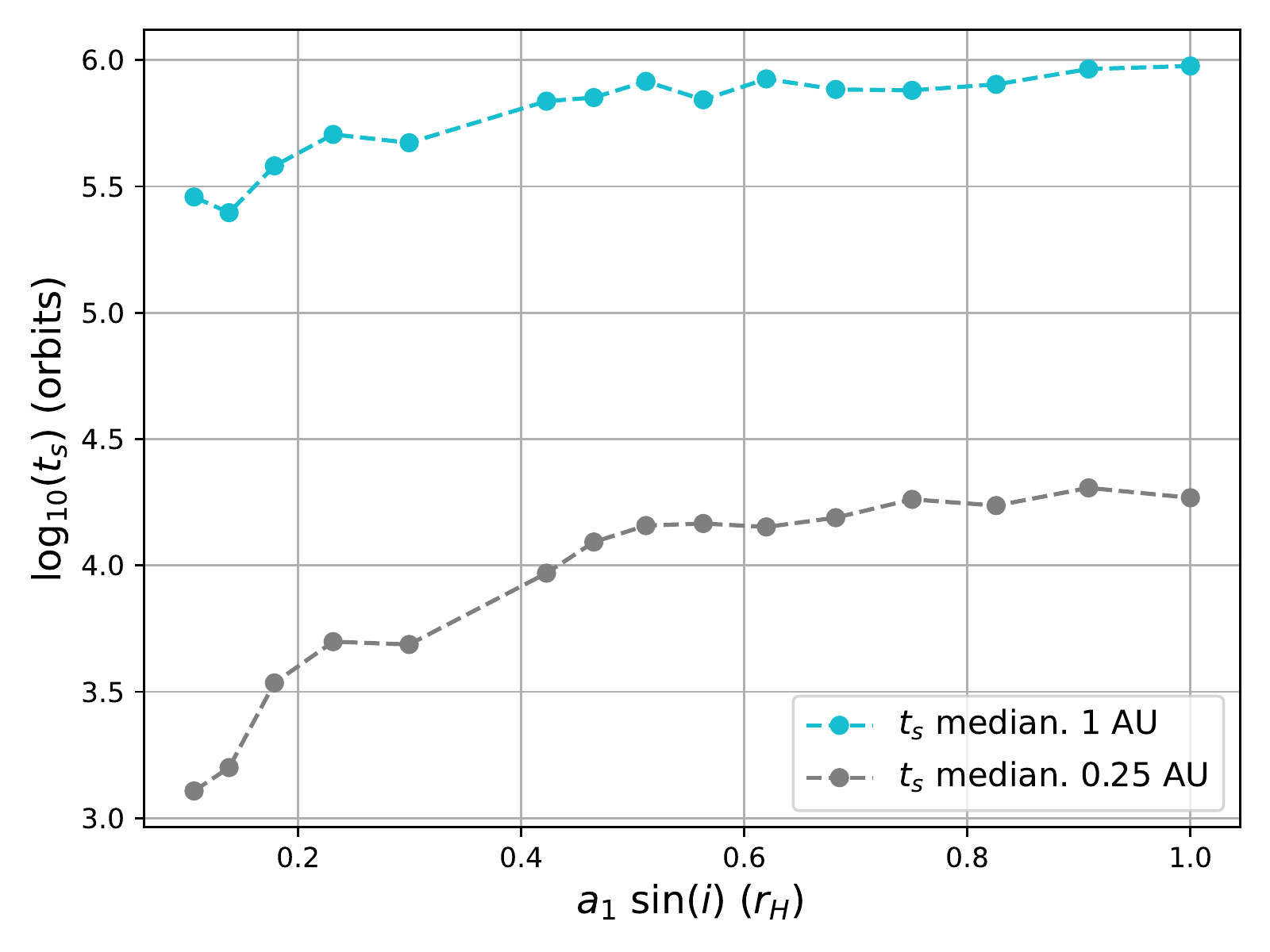}
    \caption{\rev{Median} of the log post-crossing survival time for each value of initial inclination within the inclined suite represented by the orbital height as a fraction of the Hill radius. There are fifteen values of inclination used meaning that each data point plotted is the average of up to $1120$ integrations; \rev{the only systems excluded are those that did not experience a collision in the maximum integration time}.}
    \label{fig: inclined ts vs initial orbital height}
\end{figure}

\begin{figure}
    \centering
    \includegraphics[width=0.475\textwidth]{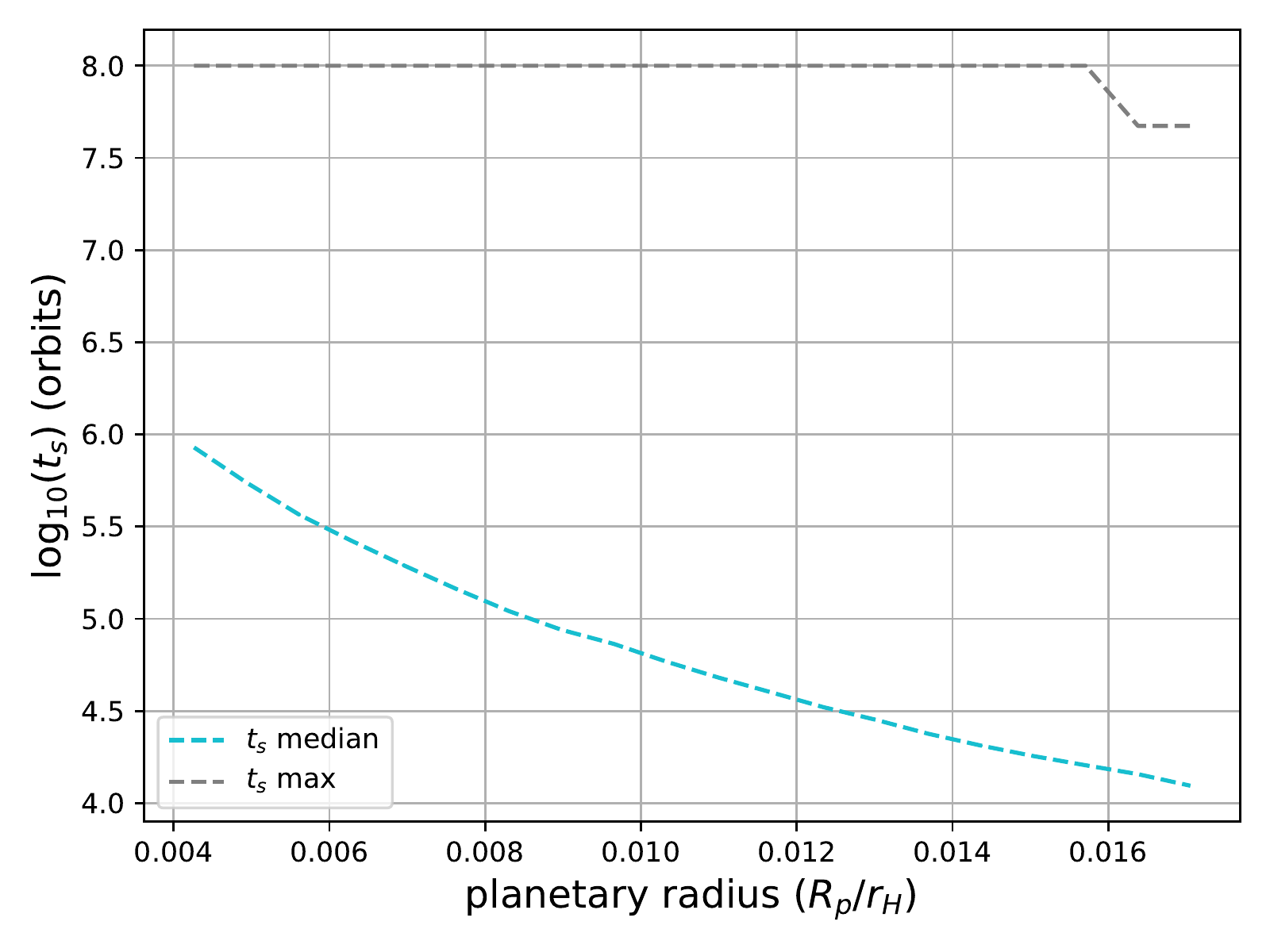}
    \caption{Median and maximum post-crossing survival time for systems as a function of the radius of planets \rev{relative to the Hill radius at $1$~AU} for systems in the inclined integration suite \rev{at $1$~AU}. Simulation times are capped at $10^8$ orbits.}
    \label{fig: planet radius vs ts}
\end{figure}

\begin{figure}
    \centering
    \includegraphics[width=0.495\textwidth]{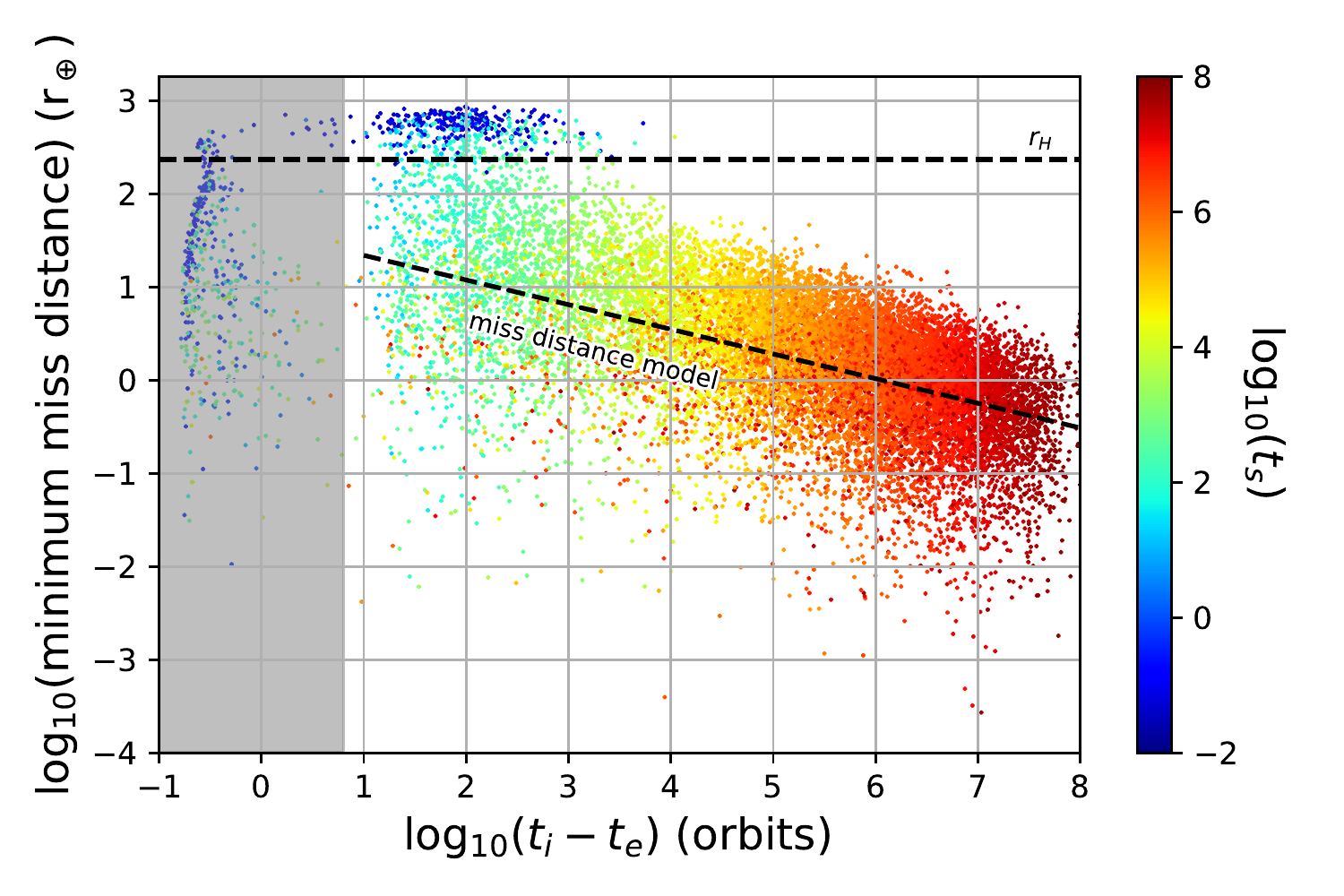}
        \caption{Time between closest encounter prior to impact and impact against the distance between the surfaces of the planets involved for systems \rev{at $1$~AU} in the inclined integration suite. The post-crossing survival time of each system is indicated through colouring. The grey shaded area indicates impacts that are possibly due to temporary gravitational capture which are excluded from the fitted model \rev{shown as a bold dashed black line}. \rev{The horizontal dashed black line shows the Hill radius at $1$~AU.}} 
    \label{fig: closest encounter}
\end{figure}

\begin{figure}
    \centering
    \includegraphics[width=0.495\textwidth]{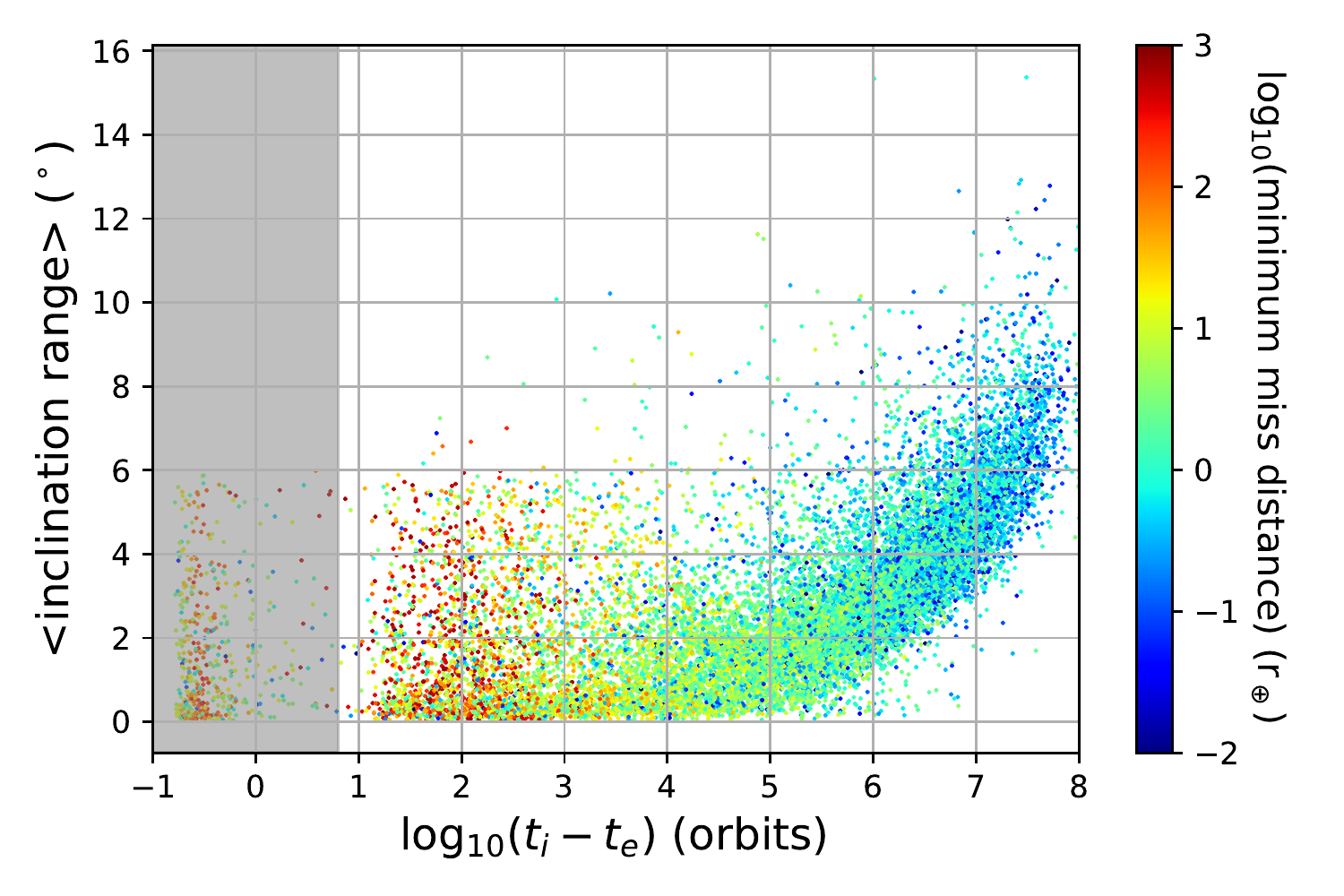}
        \caption{Time between closest encounter prior to impact and impact against the time\rev{-}averaged inclination range, i.e. the difference between the smallest and largest inclinations, for systems  \rev{at $1$~AU} in the inclined integration suite. The closest encounter experienced by a system is indicated through colouring.} 
    \label{fig: inclination vs encounter time}
\end{figure}

\begin{figure}
    \centering
    \includegraphics[width=0.495\textwidth]{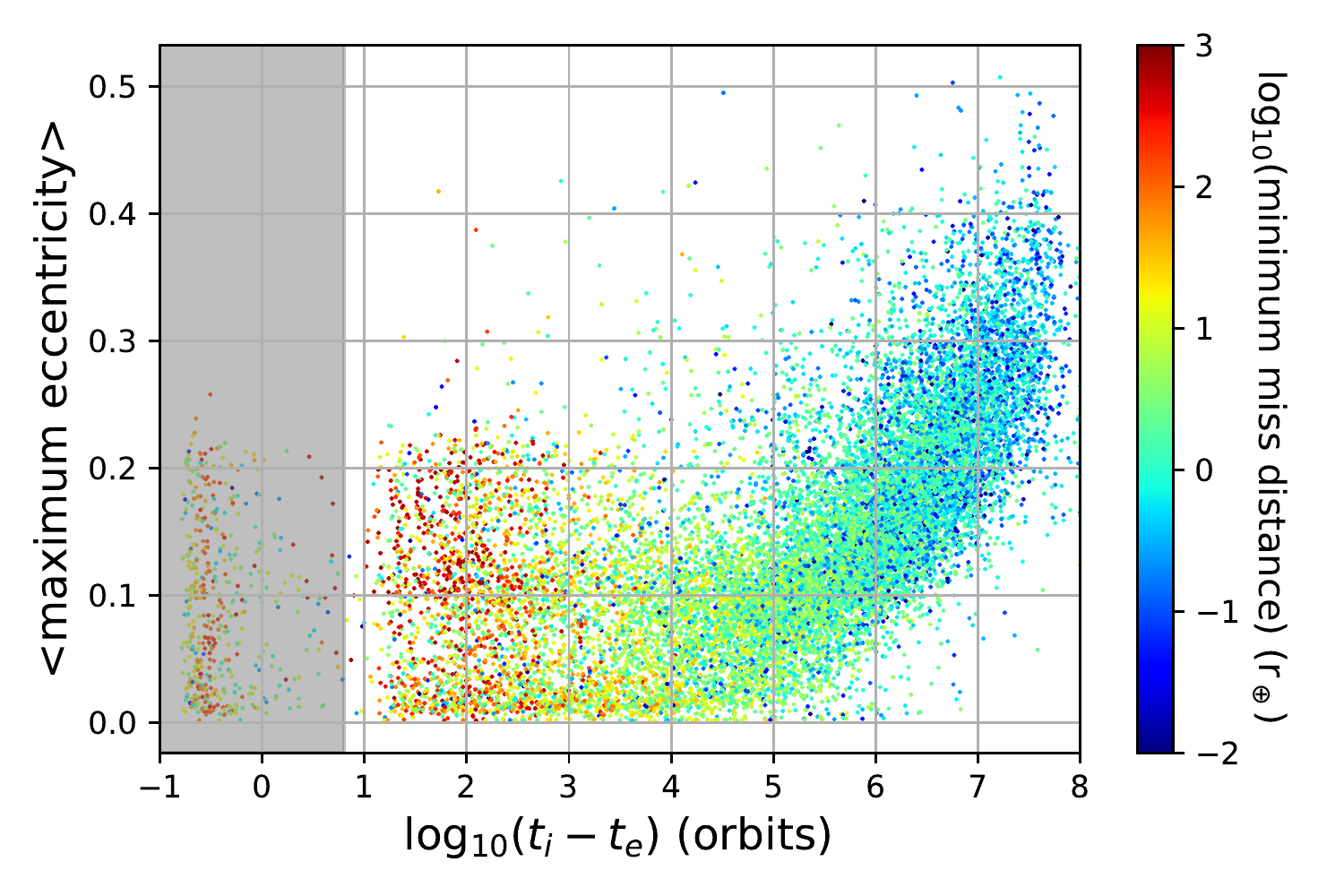}
        \caption{Time between closest encounter prior to impact and impact against the time\rev{-}averaged maximum eccentricity for systems \rev{at $1$~AU} in the inclined integration suite. The closest encounter experienced by a system is indicated through colouring.} 
    \label{fig: eccentricity vs encounter time}
\end{figure}

\subsection{Time scale to planet-planet collision}
\label{sec: inclined time scale to planet planet collision}
Figure \ref{fig: tc inclined mean min max} shows the crossing time for systems within our inclined suite. We find a large variance in crossing time across the inclined suite with a difference between the maximum and minimum crossing times at each value of $\beta$ as large as two orders of magnitude in many cases. The spikes seen in Fig.~\ref{fig: orbital crossing at 1AU} are also present in some of our inclined cases. A model of the type in Eq.~\ref{eq: model} is fitted to the mean values of crossing time observed at each value of $\beta$, yielding coefficients  $b' = \rev{1.39}$ and $c' = \rev{2.18}$. These values are in very good agreement with those from the standard suite. This is, however, where the similarities between the co-planar and inclined cases end. Figure \ref{fig: ts inclined} shows the post-crossing survival time for systems within the inclined suite, the times are much higher than in the co-planar case where the longest surviving system after crossing survived for roughly one million orbits. Here, the majority of systems survive for longer than this and, in fact, there are twenty-three systems that do not experience any collision at all within the maximum simulation time (100 million orbits), equivalent to $0.14\%$ of all integrations. Given that the post-crossing survival time is now approaching one percent of the lifetime of the Sun, it is much less unlikely that we actually could observe an inclined system between a crossing and a collision. However, at $0.25$~AU no integrations survived for the full simulation duration after an integration. 

Figure \ref{fig: ti inclined probability} shows the probability of a collision across all integrations within the inclined suite. \rev{The probability is calculated as the cumulative fraction of systems that have experienced collisions divided by the total number of systems.} Results are included for systems initially at $1$ AU as well as at $0.25$ AU. Decreasing the initial distance to the star by this amount is identical to having artificially inflated the planetary radius $\textrm{R}_p$ by a factor of four, i.e., made $\textrm{R}_p$ approximately equal to that of Neptune, \rev{whilst keeping the innermost planet initially at $1$~AU}. It is therefore expected that the collision probability over time should increase with decreasing initial distance to the star. However, the increase is striking: for Earth analogues the probability of a collision for a given system after one million orbital periods is roughly 50\%, but for a Neptune radius (1~M$_\oplus$) planet at 1 AU that probability increases to over 75\% across all $\beta$ ranges, reaching almost 90\% in all but one range.
Furthermore, for the $1$ AU systems it can be seen that the various $\beta$ regions converge after roughly a million orbits. This indicates that the evolution after the first close encounter has reconfigured the system such that any \rev{prior} collision probabilities due to initial orbital spacing are lost. To understand this, we can look at the collision probability \rev{in Figure \ref{fig: ti inclined probability} at one million orbits for $0.25$ AU systems. These systems are equivalent to to a Neptune radius planet being placed at $1$~AU and roughly $90\%$ have experienced a collision within this timescale. We can therefore infer that the same roughly $90\%$ of Earth radius planets at $1$~AU must have experienced a close encounter within $4 \textrm{R}_p$. The loss of prior collision probabilities due to orbital spacing after this point in time therefore appears to be driven by these particularly close encounters.}

Figure \ref{fig: inclined survival probabilities four panel} contains the distribution of post-crossing survival times for two subsets of the inclined suite results: the subsets each contain $1120$ configurations, one at the minimum initial inclination ($0.06^\circ$) and the other at the maximum initial inclination ($0.58^\circ$). It can be seen that the distributions are different at each initial orbital radii and inclination. Firstly, the population of collisions taking place within several orbits of an orbital crossing decreases with increasing initial inclination. In both of the most highly inclined cases there is only a single peak present in the distribution; however, this distribution is much more negatively skewed in systems initially at $1~$AU. In the lowest inclination cases there are two peaks present in addition to the one caused by immediate collisions. One \rev{peak is} collocated with those found in the more inclined case. The second peak is centered at approximately $t_s = 10^{2.5}$. In the co-planar case we have seen that the distribution of post-crossing survival times are centered at approximately $10^{2.5}$ orbits and it is also known that if the inclination is below the critical threshold $i = r_H$ the number of collisions occurring within a factor of three of the orbital crossing increases \citep{Rice2018}. Both of these factors combined explain the appearance of this second peak. Additionally, a larger proportion of systems at $0.25~$ AU experience a collision in this second peak.

The effect of increased initial inclination across the whole inclined integration suite can be seen in Fig.~\ref{fig: inclined ts vs initial orbital height}, where an increase in inclination, shown here in terms of orbital height, leads to a moderate increase in the median post-crossing survival times for \rev{systems} at \rev{both} 0.25~AU \rev{and} 1~AU. The RMS inclination in compact three-body systems has been seen to stay approximately constant up until the time of the first close encounter, which means that observed inclinations of actual planetary systems could in fact provide information about the probable survival times of systems after an orbital crossing.

The parameter that dominates the post-crossing survival time of systems in the inclined suite is the \rev{ratio of the planetary radius to the Hill radius at $1$~AU}. Figure \ref{fig: planet radius vs ts} shows the median of the log post-crossing  survival times for all systems in the suite \rev{at $1$~AU}. We find almost two orders of magnitude difference in the average survival time of systems with planets where $\nicefrac{\textrm{R}_p}{r_H} = 0.017$ as compared to systems with planets where $\nicefrac{\textrm{R}_p}{r_H} = 0.004$. This outweighs the effect of initial inclination on the survival times. Interestingly, systems surviving for the full $10^8$ orbits can be seen all the way down to a value of $\nicefrac{\textrm{R}_p}{r_H} = 0.0157$ where a rapid decrease in the lifetime of the longest lived of systems is seen. This is equivalent to a planet initially located at $1$ AU with a radius $3.5$ times that of Earth.

In addition to the dependence of the post-crossing survival time upon the orbital elements of the system, we also find a correlation with the distance of the closest approach. Figure~\ref{fig: closest encounter} shows the time taken for a collision to occur after the closest encounter experienced prior to it, at time denoted $t_e$, against the distance between the surfaces of the planets. Data points in the shaded grey area are excluded from our fitted models, and this area corresponds to the boundary seen in Figure~\ref{fig: survival time against beta with satellite images} at approximately eight orbits. Here, we see a strong negative correlation where a least squares model fitted to the log of $t_i-t_e$ and the miss distance of the encounter has a slope of $-0.26$ with a $y$-intercept of $1.6$. Ergo, the closer an encounter experienced by a system the longer it is likely to survive afterwards. In this plot, each point is also coloured according to the post-crossing survival time of the system. Looking vertically from top to bottom at the colouring it can also be seen that the absolute post-crossing survival time of systems depends upon the miss distance of the closest encounter. It seems that for planetary systems to survive for a long time after an orbital crossing they must risk collision.

We find that the closest encounters are responsible for driving the largest changes in both inclination and eccentricity, and we believe that it is the increase in inclination that causes the trend seen in Figure~\ref{fig: closest encounter}. Figure~\ref{fig: inclination vs encounter time} shows the time taken for a collision to occur after the closest encounter experienced prior to it against the time\rev{-}averaged inclination range, i.e. the difference between the maximum and minimum inclinations. Systems with the largest inclination range survive for the longest after a close encounter and the minimum miss distance, indicated through colouring, is key to increasing this range.
Figure~\ref{fig: eccentricity vs encounter time} is identical except it shows the time\rev{-}averaged maximum eccentricity in a system. Again, the minimum miss distance can be seen to be responsible for the increases in eccentricity. \rev{These increases in eccentricity will also work to increase the lifetime of systems through a reduction in the effect of gravitational focusing on the combined physical/gravitational cross-sectional area of planets \citep{Safronov1972}.}


\begin{figure}
    \centering
    \includegraphics[width=0.475\textwidth]{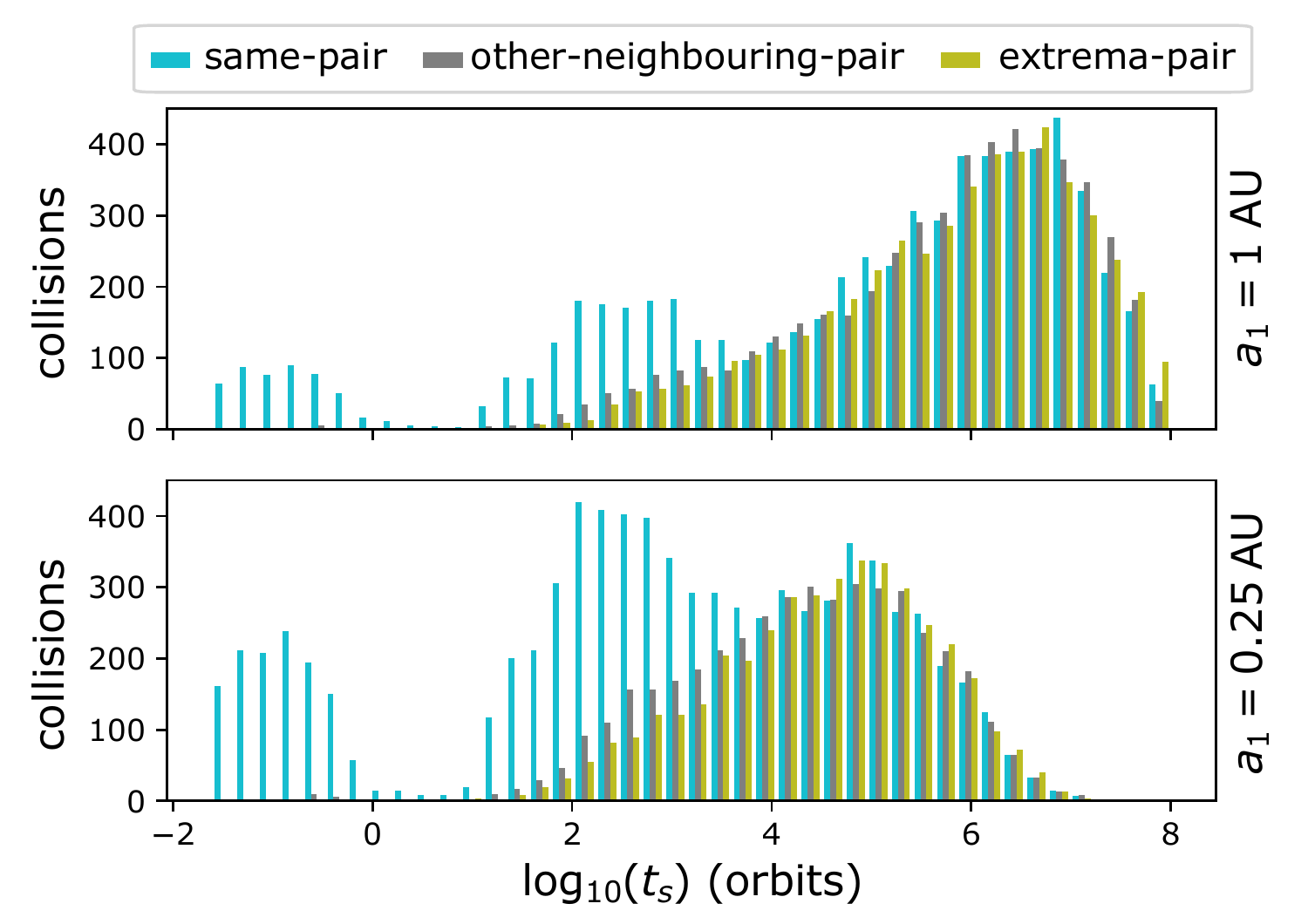}
        \caption{Time distribution of collisions between different pairs of planets in the inclined integration suite. Cyan bars indicate the same pair both crossed orbits and collided; dark grey indicates the pair that collided was not the pair that crossed; yellow indicates a collision between the inner and outer planets. The top pane\rev{l} is initially at $1$ AU while the bottom pane\rev{l} is initially at $0.25$ AU.} 
    \label{fig: inclined pairs involved in collision distribution}
\end{figure}

\subsection{Which planets collide?}

Figure \ref{fig: inclined pairs involved in collision distribution} is the equivalent to Fig. \ref{fig: pairs involved in collision distribution} but for the inclined suite. Similarly to the co-planar case, we find that collisions within a single orbit are almost exclusively between the same pair involved in the crossing. We also find an increase in the number of collisions within this time frame in the $0.25$ AU case compared to the $1$ AU case. However, at a factor of approximately $3$, here we see that the increase is more substantial.
The distributions of survival times for systems surviving after crossing for longer than a single orbit appear very different to the co-planar case. Nonetheless, some similarities in behaviour are present: in both the co-planar and inclined case there is a peak present of same-pair collisions between $10^2$ and $10^3$ orbits. Adjusting for the number of systems in each suite we find that the fraction of systems colliding at this point is roughly five times smaller at $1$ AU in the inclined case. The time period for mixing in the inclined case is approximately $10^4$ orbits, slightly longer than in the co-planar case, after which collisions between any pair of planets become equally likely.

\section{Conclusions}
\label{sec: conclusions}
We performed more than $25,000$ integrations of compact three-planet systems with the TES integration tool for a maximum time of $10^9$ orbits of the innermost planet or until the first collision of planets. We chose to focus our attention on the effects of orbital spacing and therefore distributed system configurations across a wide range of initial values evenly spaced in $\beta$. Efforts were initially focused on the co-planar case where it is easier to isolate the effects of increasing $\beta$ but then extended to include the inclined case as well.

We find in the co-planar suite that planetary systems are doomed after an orbital crossing: they rapidly experience a collision within a maximum observed time of less than one million orbits. However, despite this prognosis, we found that systems with a wider initial spacing of planets do survive longer, exhibiting a median post-crossing survival time following a slope $\textrm{log}_{10}(t_s) \propto 0.12 \, \beta$. Additionally, we show that three distinct populations of post-instability impact behaviour are present, with very few outliers:
\begin{enumerate}
    \item immediate collisions within a tenth of an orbit,
    \item prompt collisions between a tenth of an orbit and two orbits,
    \item those surviving for much longer than ten orbits.
\end{enumerate}
The pathology of these different behaviours have been identified and each of them are also observed in the inclined suite.

The probabilities of a collision between specified planetary pairs were also calculated and it was found that collisions will take place between the same pair of planets that initially crossed in the majority of cases, ranging from $48\%$ to $62\%$ depending on the region of $\beta$. These probabilities increase further depending on the radius of the planet with Neptune-radius planets experiencing probabilities as high as $76\%$. Despite this increase in probabilities in the co-planar case, the post-crossing survival time only weakly depends upon the planetary radius causing an increase of only $10^3$ orbits. In the inclined suite, however, we observe that the planetary radius is the main driver of the post-crossing survival time. We find a decrease in median post-crossing survival time of almost two orders of magnitude between Earth and Neptune radius planets. Additionally, the initial orbital inclinations have been shown to also influence the post-crossing survival times across the full range of $\beta$ by as much as an order of magnitude.

Additionally, we looked at the RMS eccentricity and inclination growth of all systems within our inclined suite after an orbital crossing. Here, we replicate the eccentricity growth rate $e \propto t^{\nicefrac{1}{6}}$ found in other studies. We do, however, find the growth rate of the inclination to be $i \propto t^{\nicefrac{1}{4}}$ instead of the $i \propto t^{\nicefrac{1}{3}}$ observed in previous work. 

Finally, we have shown that systems that experience the closest encounters also survive for the longest, and planetary systems that wish to survive must therefore live dangerously.

\section*{Acknowledgements}
We would like to acknowledge the funding provided by the Engineering and Physical Sciences Research Council (EPSRC) Centre for Doctoral Training in Next Generation Computational Modelling grant EP/L015382/1 that has made this research possible.
Additionally, we would like to acknowledge the use of the IRIDIS High Performance Computing Facility, and associated support services, at the University of Southampton.
JJL was supported through NASA's PSD ISFM program. \rev{We would like to thank an anonymous reviewer for their detailed comments that helped improve this manuscript.}

\section*{Data Availability}
We have made available all data generated from our integrations. This includes both the data files containing crossing and collision timings for all systems, and periodic state vector data. The dataset is available via \hyperlink{https://doi.org/10.5258/SOTON/D1623}{https://doi.org/10.5258/SOTON/D1623}.



\bibliographystyle{mnras}
\bibliography{stability} 



\appendix

\section{Integrator comparison}
\label{appendix: integrator comparison}
As TES is a new scheme, we have chosen to perform additional integrations making use of \rev{IAS15} such that crossing and collision times can be compared. We performed integrations using \rev{IAS15} for all runs in our standard integration suite over the range $\beta = 3.5$ to $6.3$ using the standard configuration where the tolerance parameter $\epsilon = 10^{-9}$.  Additionally, we also perform identical comparisons with the dataset of crossing times in \citet{Lissauer2021} obtained using the MVS implementation within MERCURY. \rev{Throughout this section, TES and IAS15 check for an orbital crossing on every integration step whereas the MVS and hybrid schemes check for an orbital crossing once every ten years.}
Table~\ref{tab: tes vs ias15 crossing} compares the crossing time obtained by TES and \rev{IAS15}; we find very good agreement in results especially in lower $\beta$ ranges where the lifetime of systems is shorter. In particular, for systems where $\beta < 4.0$ we find that $68\%$ of systems experience an orbital crossing within $1\%$ of one another, increasing to $79\%$ if the tolerance is relaxed to being within $10\%$ of one another. These percentages can be compared to the data presented in Table~\ref{tab:crossing table} comparing the performance of TES under the influence of an initial $100$~m perturbation in the position of the innermost planet along its orbital arc. Comparison of the summary columns in these two tables reveals that the difference in crossing time between TES and \rev{IAS15} is smaller than the effect of the $100$~m perturbation.
Tables~\ref{tab: tes vs mvs crossing} and \ref{tab: ias15 vs mvs crossing} compare the crossing times found in TES and \rev{IAS15}, respectively, against th\rev{ose} obtained with MERCURY. Unsurprisingly, we find that the comparison yields very similar statistics in both cases. In particular, the runs finishing within $1\%$ and $10\%$ of each other drop by at least $54\%$ in the lowest region of $\beta$, although the reduction is much less pronounced at higher $\beta$. We also find that the difference in crossing times between TES and \rev{IAS15} at higher $\beta$ is very similar to that of TES and \rev{IAS15}, and MVS.
Finally, Table~\ref{tab: tes vs ias15 collision} compares the collision times for TES and \rev{IAS15}, and this is where we find the largest differences between the two schemes. In the summary column, over the entire $\beta$ range, we see that the number of runs finishing within $1\%$, $10\%$ and a factor of two have decreased substantially when compared to Table~\ref{tab: tes vs ias15 crossing} performing the same comparison but for crossing times. The majority of the reduction in these statistics comes from the integrations where $\beta < 5$ and the evolution after crossing is still a substantial fraction of the overall simulation time. These differences highlight the sensitivity of integrations to close encounters.

\begin{figure}
    \centering
    \includegraphics[width=0.475\textwidth]{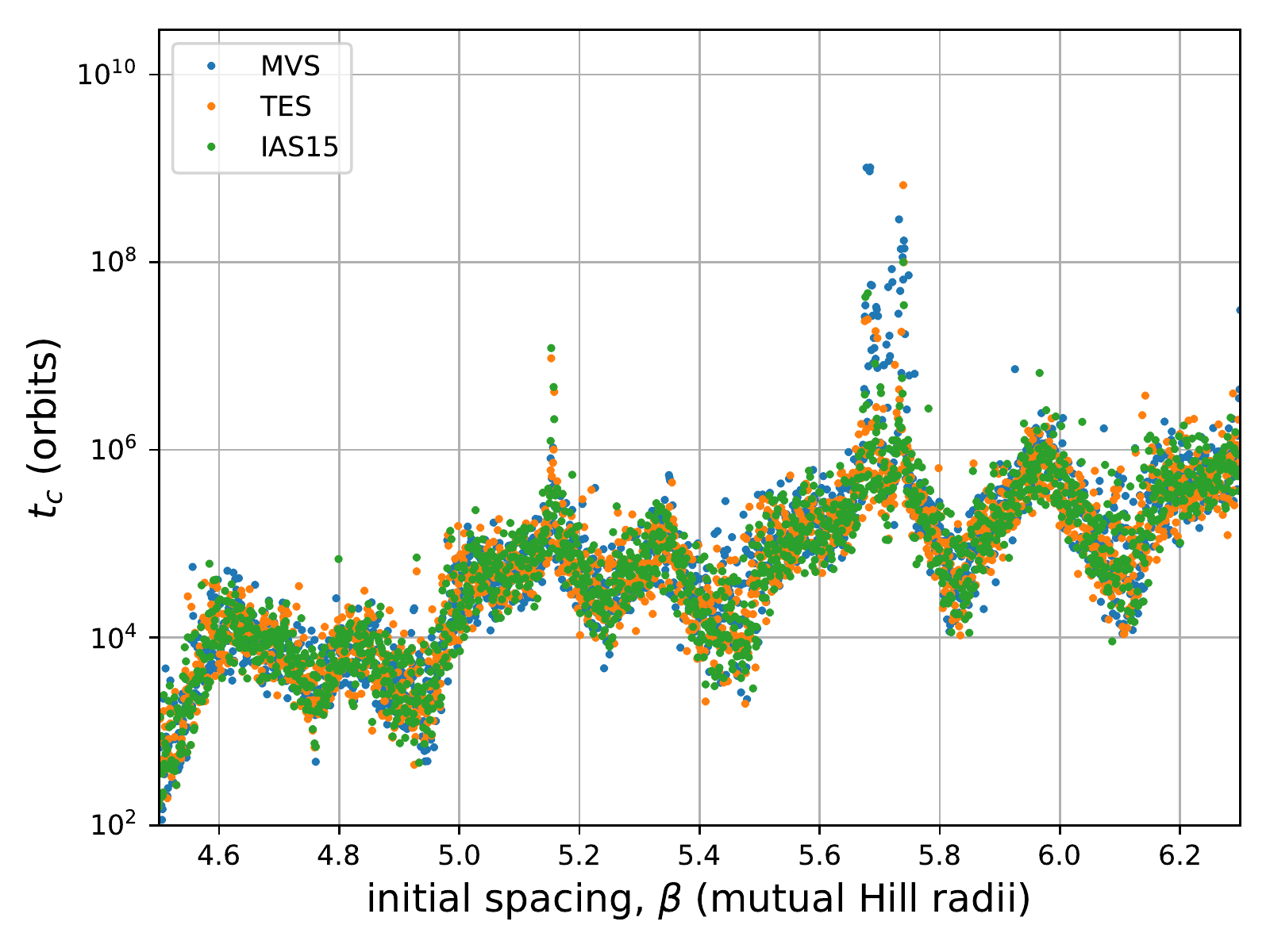}
        \caption{Plot showing a comparison of crossing times for three integration routines making use of the initial conditions in the standard integration suite.} 
    \label{fig: integrator comparison without hybrid}
\end{figure}

\begin{figure}
    \centering
    \includegraphics[width=0.475\textwidth]{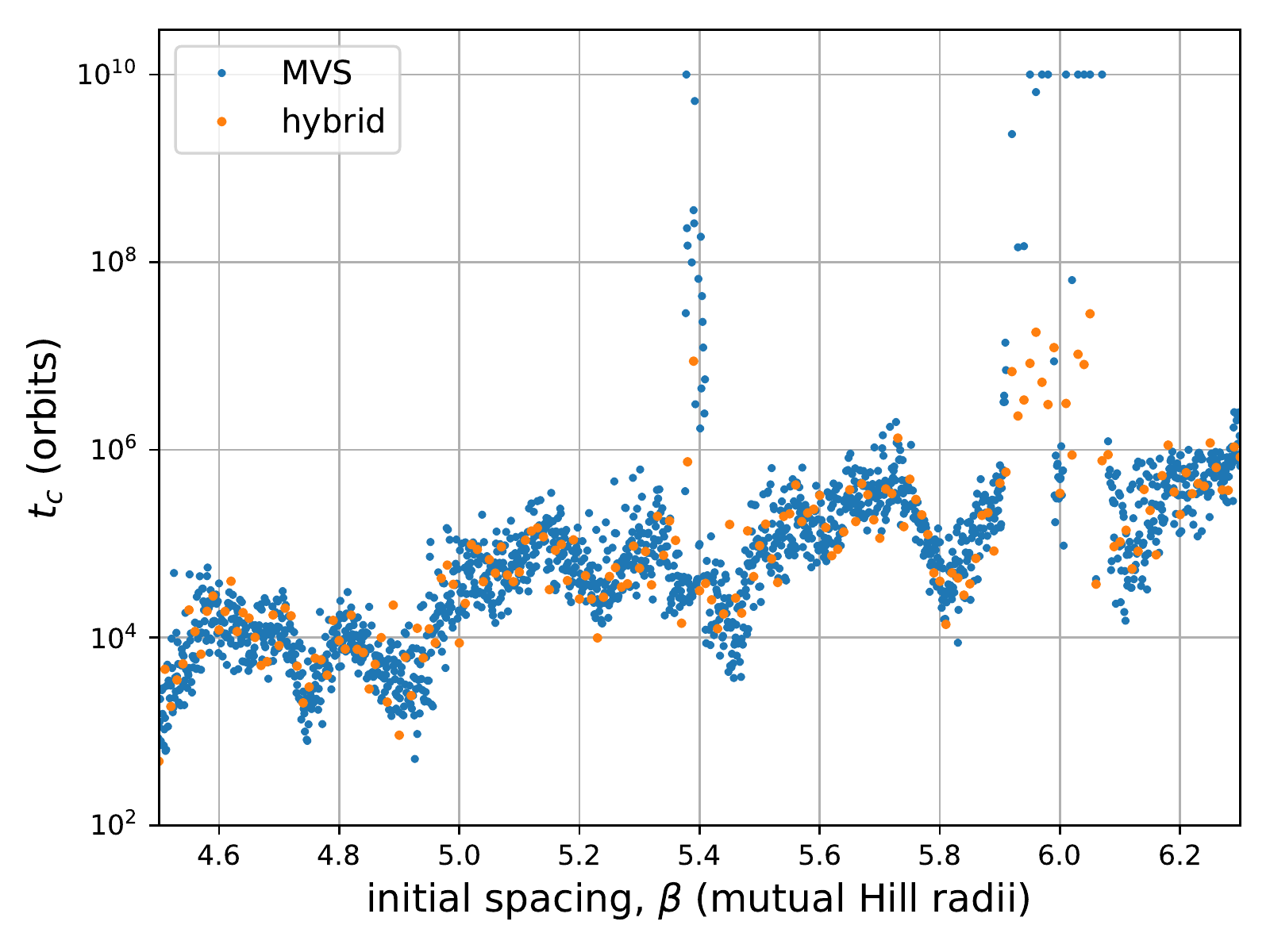}
        \caption{Plot showing a comparison of crossing time for two integrations routines with shifted initial longitudes described in the main body of text. \rev{The MVS and hybrid schemes have a density of one thousand and one hundred runs per unit $\beta$, respectively. }} 
    \label{fig: integrator comparison with hybrid}
\end{figure}

In addition to the quantitative comparisons between integrators thus far we have also encountered some qualitative differences in behaviour between the schemes examined so far and the hybrid integration scheme within MERCURY. Figure~\ref{fig: integrator comparison without hybrid} visualises the comparison between integration schemes found in Table~\ref{tab: tes vs ias15 crossing} to \ref{tab: ias15 vs mvs crossing}. It shows how tightly the crossing times for the three schemes are clustered to one another, and in particular it shows that in the region located around $\beta=5.7$, where a region of high stability is found, the schemes all perform identically capturing the long-lived system behaviour. The hybrid integration scheme within MERCURY combines the MVS scheme with a non-symplectic integrator to allow for close approaches to be handled, and it would therefore seem like an ideal candidate for performing all of the experiments in this article. Figure~\ref{fig: integrator comparison with hybrid} shows the results of an experiment identical to that described in Section~\ref{sec: standard suite} except with initial longitudes of $M_j = [0, 10.17, 20.33]^{\circ}$. \rev{In this experiment, the MVS scheme has a density of one thousand integrations per unit $\beta$ whereas the hybrid scheme has a reduced density of one hundred per unit $\beta$.} Both the MVS and hybrid schemes use what is considered a conservative step size of $18$ days resulting in slightly over $20$ steps per orbit. For the most part, the schemes agree well with each other; however, a key difference between the schemes can be seen in the region $\beta=6$ where no integrations performed by the hybrid scheme lasted for longer than $10^8$ orbits despite the majority of MVS scheme integrations lasting for $10^{10}$ orbits. It appears that the hybrid scheme is not accurate enough properly to capture the dynamics in this region which has led to a population of short\rev{-}lived systems that in fact should have been stable for a lot longer. This was a strong motivation for not using the hybrid scheme for experiments in this work instead opting to use TES which, as can be seen in Fig.~\ref{fig: integrator comparison without hybrid}, can properly capture the dynamics in regions of increased stability.

\begin{table*}
	\centering
	\caption{Comparison of \emph{crossing times} of systems using identical values of $\beta$ for the standard initial longitudes using \emph{TES} and \emph{\rev{IAS15}}. Systems have the innermost planet initially placed at $1$ AU.}
	\begin{tabular}{lcccccc} 
		\hline
		Interval: & [3.5, 3.999] & [4.0, 4.999] &  [5.0, 5.999] & [6.0, 6.3] &  \hspace{0.3cm} & [3.5, 6.3]\\
        \hline
        number of runs in the range & $500$  & $1000$  & $1000$  & $301$  && $2801$  \\
        $< \textrm{log}_{t_c}\textrm{(TES)} - \textrm{log}_{t_c}\textrm{\rev{(IAS15)}}>$ & $-0.014$  & $-0.005$  & $0.004$  & $-0.041$  && $-0.007$  \\
        $< |\textrm{log}_{t_c}\textrm{(TES)} - \textrm{log}_{t_c}\textrm{\rev{(IAS15)}} |>$ & $0.034$  & $0.156$  & $0.289$  & $0.352$  && $0.203$  \\
        $t_c$\rev{(IAS15)} < $0.5 t_c$(TES) & $2$ ($0.40\%$)  & $77$ ($7.70\%$)  & $200$ ($20.00\%$)  & $56$ ($18.60\%$)  && $335$ ($11.96\%$)  \\
        $0.5 t_c$(TES) < $t_c$\rev{(IAS15)} < $2 t_c$(TES) & $489$ ($97.80\%$)  & $848$ ($84.80\%$)  & $608$ ($60.80\%$)  & $164$ ($54.49\%$)  && $2109$ ($75.29\%$)  \\
        $t_c$(TES) < $0.5 t_c$\rev{(IAS15)} & $9$ ($1.80\%$)  & $75$ ($7.50\%$)  & $192$ ($19.20\%$)  & $81$ ($26.91\%$)  && $357$ ($12.75\%$)  \\
        within 10\% of one another & $395$ ($79.00\%$)  & $291$ ($29.10\%$)  & $105$ ($10.50\%$)  & $27$ ($8.97\%$)  && $818$ ($29.20\%$)  \\
        within 1\% of one another & $340$ ($68.00\%$)  & $120$ ($12.00\%$)  & $9$ ($0.90\%$)  & $1$ ($0.33\%$)  && $470$ ($16.78\%$)  \\
        \hline
	\end{tabular}
	\label{tab: tes vs ias15 crossing}
\end{table*}

\begin{table*}
	\centering
	\caption{Comparison of \emph{crossing times} of systems using identical values of $\beta$ for the standard initial longitudes using \emph{TES} and \emph{MVS} with the innermost planet initially at $1$ AU. \rev{Data marked with a * are likely to be somewhat erroneous due to the MVS scheme integrations only checking for an orbital crossing once every ten orbits.}}
	\begin{tabular}{lcccccc} 
		\hline
		Interval: & [3.5, 3.999] & [4.0, 4.999] &  [5.0, 5.999] & [6.0, 6.3] &  \hspace{0.3cm} & [3.5, 6.3]\\
        \hline
        number of runs in the range & $500$  & $1000$  & $1000$  & $301$  && $2801$  \\
        $< \textrm{log}_{t_c}\textrm{(TES)} - \textrm{log}_{t_c}\textrm{(MVS)}>$ & $-0.045$  & $0.002$  & $-0.081$  & $-0.044$  && $-0.041$  \\
        $< |\textrm{log}_{t_c}\textrm{(TES)} - \textrm{log}_{t_c}\textrm{(MVS)} |>$ & $0.244$  & $0.303$  & $0.355$  & $0.371$  && $0.318$  \\
        $t_c$(MVS) < $0.5 t_c$(TES) & $64$ ($12.80\%$)  & $215$ ($21.50\%$)  & $172$ ($17.20\%$)  & $56$ ($18.60\%$)  && $507$ ($18.10\%$)  \\
        $0.5 t_c$(TES) < $t_c$(MVS) < $2 t_c$(TES) & $335$ ($67.00\%$)  & $589$ ($58.90\%$)  & $558$ ($55.80\%$)  & $162$ ($53.82\%$)  && $1644$ ($58.69\%$)  \\
        $t_c$(TES) < $0.5 t_c$(MVS) & $101$ ($20.20\%$)  & $196$ ($19.60\%$)  & $270$ ($27.00\%$)  & $83$ ($27.57\%$)  && $650$ ($23.21\%$)  \\
        within 10\% of one another & $67$ ($13.40\%$)  & $99$ ($9.90\%$)  & $81$ ($8.10\%$)  & $21$ ($6.98\%$)  && $268$ ($9.57\%$)  \\
        within 1\% of one another & $12$ ($2.40\%$)\rev{*}  & $4$ ($0.40\%$)\rev{*}  & $6$ ($0.60\%$)  & $4$ ($1.33\%$)  && $26$ ($0.93\%$)\rev{*}  \\
        \hline
	\end{tabular}
	\label{tab: tes vs mvs crossing}
\end{table*}

\begin{table*}
	\centering
	\caption{Comparison of \emph{crossing times} of systems using identical values of $\beta$ for the standard initial longitudes using \emph{\rev{IAS15}} and \emph{MVS} with the innermost planet initially at $1$ AU. \rev{Data marked with a * are likely to be somewhat erroneous due to the MVS scheme integrations only checking for an orbital crossing once every ten orbits.}}
	\begin{tabular}{lcccccc} 
		\hline
		Interval: & [3.5, 3.999] & [4.0, 4.999] &  [5.0, 5.999] & [6.0, 6.3] &  \hspace{0.3cm} & [3.5, 6.3]\\
        \hline
        number of runs in the range & $500$  & $1000$  & $1000$  & $301$  && $2801$  \\
        $< \textrm{log}_{t_c}\textrm{\rev{(IAS15)}} - \textrm{log}_{t_c}\textrm{(MVS)}>$ & $-0.031$  & $0.008$  & $-0.086$  & $-0.002$  && $-0.034$  \\
        $< |\textrm{log}_{t_c}\textrm{\rev{(IAS15)}} - \textrm{log}_{t_c}\textrm{(MVS)} |>$ & $0.243$  & $0.291$  & $0.359$  & $0.360$  && $0.314$  \\
        $t_c$(MVS) < $0.5 t_c$\rev{(IAS15)} & $72$ ($14.40\%$)  & $201$ ($20.10\%$)  & $189$ ($18.90\%$)  & $74$ ($24.58\%$)  && $536$ ($19.14\%$)  \\
        $0.5 t_c$\rev{(IAS15)} < $t_c$(MVS) < $2 t_c$\rev{(IAS15)} & $333$ ($66.60\%$)  & $605$ ($60.50\%$)  & $552$ ($55.20\%$)  & $155$ ($51.50\%$)  && $1645$ ($58.73\%$)  \\
        $t_c$\rev{(IAS15)} < $0.5 t_c$(MVS) & $95$ ($19.00\%$)  & $194$ ($19.40\%$)  & $259$ ($25.90\%$)  & $72$ ($23.92\%$)  && $620$ ($22.13\%$)  \\
        within 10\% of one another & $66$ ($13.20\%$)  & $112$ ($11.20\%$)  & $79$ ($7.90\%$)  & $27$ ($8.97\%$)  && $284$ ($10.14\%$)  \\
        within 1\% of one another & $10$ ($2.00\%$)\rev{*}  & $5$ ($0.50\%$)\rev{*}  & $10$ ($1.00\%$)  & $2$ ($0.66\%$)  && $27$ ($0.96\%$)\rev{*}  \\
        \hline
	\end{tabular}
	\label{tab: ias15 vs mvs crossing}
\end{table*}

\begin{table*}
	\centering
	\caption{Comparison of \emph{collision times} of systems using identical values of $\beta$ for the standard initial longitudes using \emph{TES} and \emph{\rev{IAS15}} with the innermost planet initially at $1$ AU.}
	\begin{tabular}{lcccccc} 
		\hline
		Interval: & [3.5, 3.999] & [4.0, 4.999] &  [5.0, 5.999] & [6.0, 6.3] &  \hspace{0.3cm} & [3.5, 6.3]\\
        \hline
        number of runs in the range & $500$  & $1000$  & $1000$  & $301$  && $2801$  \\
        $< \textrm{log}_{t_c}\textrm{(TES)} - \textrm{log}_{t_c}\textrm{\rev{(IAS15)}}>$ & $-0.080$  & $-0.024$  & $0.003$  & $-0.039$  && $-0.026$  \\
        $< |\textrm{log}_{t_c}\textrm{(TES)} - \textrm{log}_{t_c}\textrm{\rev{(IAS15)}} |>$ & $0.485$  & $0.296$  & $0.280$  & $0.352$  && $0.330$  \\
        $t_c$\rev{(IAS15)} < $0.5 t_c$(TES) & $125$ ($25.00\%$)  & $170$ ($17.00\%$)  & $189$ ($18.90\%$)  & $59$ ($19.60\%$)  && $543$ ($19.39\%$)  \\
        $0.5 t_c$(TES) < $t_c$\rev{(IAS15)} < $2 t_c$(TES) & $207$ ($41.40\%$)  & $626$ ($62.60\%$)  & $622$ ($62.20\%$)  & $156$ ($51.83\%$)  && $1611$ ($57.52\%$)  \\
        $t_c$(TES) < $0.5 t_c$\rev{(IAS15)} & $168$ ($33.60\%$)  & $204$ ($20.40\%$)  & $189$ ($18.90\%$)  & $86$ ($28.57\%$)  && $647$ ($23.10\%$)  \\
        within 10\% of one another & $62$ ($12.40\%$)  & $117$ ($11.70\%$)  & $115$ ($11.50\%$)  & $29$ ($9.63\%$)  && $323$ ($11.53\%$)  \\
        within 1\% of one another & $33$ ($6.60\%$)  & $20$ ($2.00\%$)  & $14$ ($1.40\%$)  & $1$ ($0.33\%$)  && $68$ ($2.43\%$)  \\
        \hline
	\end{tabular}
	\label{tab: tes vs ias15 collision}
\end{table*}


\bsp	
\label{lastpage}
\end{document}